\documentclass[journal]{vgtc}        

\onlineid{1609}

\vgtccategory{Research}

\vgtcpapertype{Applications}

\title{TROPHY: A Topologically Robust Physics-Informed Tracking Framework for Tropical Cyclones}

\author{
  \authororcid{Lin Yan}{0000-0001-7017-0329},
  \authororcid{Hanqi Guo}{0000-0001-7776-1834}, \authororcid{Thomas Peterka}{0000-0002-0525-3205}, \authororcid{Bei Wang}{0000-0002-9240-0700}, and 
  \authororcid{Jiali Wang}{0000-0002-8916-4372}
}

\authorfooter{
  \item
  	Lin Yan is with Argonne National Laboratory.
  	E-mail: lyan@anl.gov.
  \item
  	Hanqi Guo is with the Ohio State University.
  	E-mail: guo.2154@osu.edu.
  \item Thomas Peterka is with Argonne National Laboratory.
  	E-mail: tpeterka@mcs.anl.gov.
  \item Bei Wang is with the University of Utah.
  	E-mail: beiwang@sci.utah.edu.
   \item Jiali Wang is with Argonne National Laboratory.
   E-mail: jialiwang@anl.gov.
}

\abstract{
Tropical cyclones (TCs) are among the most destructive weather systems. 
Realistically and efficiently detecting and tracking TCs are critical for assessing their impacts and risks. 
In particular, the eye is a signature feature of a mature TC. 
Therefore, knowing the eyes' locations and movements  is crucial for both operational weather forecasts and climate risk assessments. 
Recently, a multilevel robustness framework has been introduced to study the critical points of time-varying vector fields. 
The framework quantifies the robustness (i.e., structural stability) of critical points across varying neighborhoods. 
By relating the multilevel robustness with critical point tracking, the framework has demonstrated its potential in cyclone tracking. 
An advantage is that it identifies cyclonic features using only 2D wind vector fields, which is encouraging as most tracking algorithms require multiple dynamic and thermodynamic variables at different altitudes. 
A disadvantage is that the framework does not scale well computationally for datasets containing a large number of cyclones.  
This paper introduces a topologically robust physics-informed tracking framework (TROPHY) for TC tracking. 
The main idea is to integrate physical knowledge of TC to drastically improve the computational efficiency of  multilevel robustness framework for large-scale  climate datasets. 
First, during preprocessing, we propose a physics-informed feature selection strategy to filter $90\%$ of critical points that are short-lived and have low stability, thus preserving good candidates for TC tracking. 
Second, during in-processing, we impose constraints during the multilevel robustness computation to focus only on physics-informed neighborhoods of TCs. 
We apply TROPHY to 30 years of 2D wind fields from reanalysis data in ERA5 and generate a number of TC tracks. 
In comparison with the observed tracks, we demonstrate that TROPHY can capture TC characteristics (e.g., frequency, intensity, duration, latitudes with maximum intensity, and genesis) that are comparable to and sometimes even better than a well-validated TC tracking algorithm that requires multiple dynamic and thermodynamic scalar fields.
}


\graphicspath{{figs/}} 

\usepackage{tabu}                      
\usepackage{booktabs}                  
\usepackage{multirow}
\usepackage{amssymb}
\usepackage{amsmath} 
\usepackage{amsfonts}

\usepackage{mathtools}

\newcommand{\para}[1]        {\vspace{1pt}\noindent{\textbf{#1}}}

\newcommand {\mm}[1] {\ifmmode{#1}\else{\mbox{\(#1\)}}\fi}

\newcommand{\Xspace}        {\mm{\mathbb{X}}}
\newcommand{\Rspace}        {\mm{\mathbb{R}}}

\newcommand{\EW}        {\mm{\mathsf{ERA5\,Wind}}}
\newcommand{\EY}        {\mm{\mathsf{ERA5\,Year}}}
\newcommand{\EEight}        {\mm{\mathsf{ERA5\,1998}}}
\newcommand{\EFour}        {\mm{\mathsf{ERA5\,2004}}}
\newcommand{\ETen}        {\mm{\mathsf{ERA5\,2010}}}
\newcommand{\TCD}        {\mm{\mathrm{TCD}}}
\newcommand{\ocr}        {\mm{\mathrm{ocr}}}

\newcommand{\minR}        {\mm{\mathrm{minR}}}

\newcommand{\mydeg}{\mm{\mathrm{deg}}}

\newcommand{\etal}{{et al.}}
\newcommand{\eg}{{e.g.}}
\newcommand{\ie}{{i.e.}}
\newcommand{\wrt}{{w.r.t.}}

\newcommand{\TE}{{TempestExtremes}}

\newcommand{\tool}{{TROPHY}}
\newcommand{\MR}{{MRT}}

\newcommand{\myedit}[1]{{\textcolor{black}{#1}}}

\begin{document}

\firstsection{Introduction}
\maketitle

Tropical cyclones (TCs) are the largest drivers of losses among natural hazards, bringing wind gusts, high waves, storm surges, and heavy rainfall. In order to achieve improved forecasts, robust risk assessment, and confident future projections of TCs, realistically detecting and tracking TCs are critical \cite{walsh2016tropical, marchok2021important, bourdin2022intercomparison}. In particular, the eye (a region of mostly calm weather at the center of TCs) is a signature feature of a mature TC. 
Therefore, knowing the location and the movement of the eye precisely and in a timely manner is crucial for weather centers issuing warnings to the general public \cite{wong2008automatic,chang2009algorithm}. Potential forecast track errors, due to forecasting models or tracker issues, could  have negative impacts on downstream applications, such as the selection of areas under watch and warnings, and false input for storm surge models over certain coastal regions~\cite{TaylorGlahn2008}. 
In addition, since the 1980s, there have been  increasing trends in TC intensities (particularly in strong categories, such as Categories 4 and 5 hurricanes) based on the long-term observation data in the Atlantic basin~\cite{HollandBruyere2014,KossinOlanderKnapp2013}.

In the past decades, many TC trackers were developed by research institutes and weather forecast centers~\cite{CamargoZebiak2002,KleppekMuccioneRaible2008, UllrichZarzycki2017, HodgesCobbVidale2017, BellChandTory2018}. 
Most of them focused on tracking the TC regions instead of TC eyes, and they used different weather variables (e.g., minimum sea level pressure or maximum vorticity) as the basis to identify a TC candidate before applying various thresholds~\cite{HodgesCobbVidale2017, ZarzyckiUllrich2017,UllrichZarzycki2017}. 
Some efforts focused on fixing the TC eyes during tracking (e.g.,~\cite{SivaramakrishnanSelvam1966, Kepert2005, LeeMarks2000, Willoughby1998,Blackwell2000}). 

Vector field topology has seen widespread applications in science and engineering since its introduction to visualization more than 30 years ago~\cite{HelmanHesselink1989}, including climate study and ocean modeling~\cite{EngelkeMasoodBeren2020,NilssonEngelkeFriederici2020,DoraiswamyNatarajanNanjundiah2013}. 
It has been one of the most promising tools to describe and interpret vector field behaviors by providing meaningful abstraction and summarization, especially for large-scale scientific data~\cite{BujackYanHotz2020}. 
Critical points (\ie, where a vector field vanishes) are core features of vector field topology, and they can be used for studying TCs, since TC eyes can naturally be identified as critical points of the wind vector fields. 
Thus, the tracking of TCs can be converted to critical point tracking in vector field topology. 

Many algorithms have been developed to find the correspondences between critical points in successive time steps in the form of  tracks (trajectories). 
Most critical point tracking algorithms infer correspondences between critical points based on distance proximity~\cite{HelmanHesselink1990, ReininghausKastenWeinkauf2012, GuoLenzXu2021}, which may produce artifacts in TC tracking.
Wang~\etal~\cite{WangRosenSkraba2013} introduced a topological notion of robustness  to quantify the structural stability of critical points. The robustness of a critical point is the minimum amount of perturbation to the vector field necessary to cancel it. 
Skraba and Wang~\cite{SkrabaWang2014} established the theoretical foundation to relate critical point tracking with robustness: critical points with high robustness values could be tracked more easily and more accurately. 

Recently, Yan~\etal~\cite{YanUllrichVan-Roekel2022} brought this theory to practice by introducing a multilevel robustness framework for the study of 2D time-varying vector fields, which has demonstrated its potential in cyclone tracking (this is referred to as the {\MR} framework for comparison purpose).  
The multilevel robustness can be integrated with state-of-the-art feature-tracking algorithms, such as the Feature Tracking Kit (FTK)~\cite{GuoLenzXu2021}, to improve tracking results. 
An advantage is that it identifies cyclonic features using only 2D wind vector fields, which is encouraging as most TC tracking algorithms require multiple dynamic and thermodynamic variables at different altitudes. A disadvantage is that the framework does not scale well for datasets containing a large number of cyclones. 

\para{Contributions.} 
We introduce a topologically robust physics-informed tracking framework ({\tool}) for TC tracking. 
The main idea is to integrate physical knowledge of TC to drastically improve the computational efficiency of the multilevel robustness framework for large-scale climate datasets. 
Our newly designed framework, {\tool}, inherits the capability of critical point tracking based on multilevel robustness~\cite{YanUllrichVan-Roekel2022}. 
First, {\tool} tracks the TC eyes instead of the TC impact areas. 
Second, it is super lightweight, requiring only near-surface wind speeds and directions. 
Third, it is able to work with any TC/critical point tracking algorithms.  
In particular, {\tool} is customized by adding a number of physics-informed strategies on top of the {\MR} framework, making TC tracking more efficient and accurate for large-scale climate datasets. 
Our contributions are three-fold. 

First, we introduce a physics-informed feature selection strategy to filter short-lived and unstable features. Such a strategy removes $90\%$ of critical points in the multilevel robustness computation and makes {\tool} much more efficient than the previous approach~\cite{YanUllrichVan-Roekel2022}.

Second, we propose an adaptive strategy to use physics-informed local neighborhoods for the multilevel robustness computation, making it more efficient and physically meaningful under the real-world scenario. 

Third, we apply {\tool} to 30 years of reanalysis data from ERA5. 
We demonstrate that {\tool} can achieve TC tracking results comparable to and sometimes even better than those of the traditionally well-validated TC tracking algorithm,  TempestExtemes~\cite{ZarzyckiUllrich2017,UllrichZarzycki2017}.
These experimental results are encouraging since {\tool} only requires 2D wind vector field data at the near-surface, whereas the traditional TC tracking algorithms need far more variables at various altitudes.
As pointed out by Bujack~\etal~\cite{BujackYanHotz2020}, it is difficult to interpret flow topology~\wrt~physical meaning in the time-varying setting.
Our comparison between {\tool} and TempestExtemes builds a bridge between tracking methods based on vector field topology and those based on multivariate scalar fields, which helps increase the physical interpretability of vector field topology.

%\jiali{Lin, if you have time, can you read this Bourdin paper and see whether they mention any other ERA5 problems regarding TCs~\cite{walsh2016tropical, marchok2021important, bourdin2022intercomparison}}

\section{Related Work}
\label{sec:related}

We review related work on traditional TC tracking algorithms and critical point tracking from vector field topology.

\subsection{Tropical Cyclone Tracking Algorithms}
TC tracking has been studied over the past decades for weather forecasting and climate analysis to issue early warnings, assess impacted areas, and provide risk assessments for public and critical infrastructures.
Most of the \myedit{traditional} TC tracking algorithms require multiple dynamic and thermodynamic variables at different altitudes to detect and track a TC. These algorithms determine a detected feature as a TC candidate by tuning parameter thresholds; however, the choices of thresholds are generally subjective~\cite{EnzEngelmannLohmann2022}.
An example of a traditional TC tracking algorithms is TRACK~\cite{HodgesCobbVidale2017}, which uses relative vorticity at 850, 700, 600, 500, and 250 hPa as the key variable. 
TRACK uses certain criteria (still based on vorticity) during post-tracking to isolate the warm core of TCs. 
The criteria must be jointly attained for at least one day. 
Another example of \myedit{a traditional TC tracker }is {\TE}~\cite{ZarzyckiUllrich2017,UllrichZarzycki2017}, which is used in this study for comparison with {\tool} in~\cref{sec:results}. 
{\TE} uses sea level pressure (SLP) as its key feature-tracking variable. 
TC candidates are initially identified by minima and a closed contour of the SLP field. Next, a geopotential height difference between 250 to 500 hPa is used to refine the candidate definitions. 
Instead of the living time, {\TE}~requires that a tracked storm travels at least $8^\circ$ between 10 N to 40 N latitudes. 
Both TRACK and {\TE} algorithms and many others (\eg,~\cite{biswas2018hurricane}) require many input variables that are often not readily available from raw model output and need additional calculations, which can involve handling a big amount of data. 
Also, these algorithms are not exactly tracking the eyes of TCs. They typically identify minimum SLP and maximum vorticity, where the tracked path of a TC is not exactly along the eye of the TC, and sometimes the path can even be biased toward the eyewall (a ring of tall thunderstorms that produce heavy rains and usually the strongest winds).
%, which has significantly different wind speeds. 

\myedit{Recently, several contour-based and topology-based feature tracking approaches have been proposed~\cite{WischgollScheuermann2001, EdelsbrunnerHarerMascarenhas2004, SohnBajaj2005, ReininghausKastenWeinkauf2012, DoraiswamyNatarajanNanjundiah2013, SkrabaWang2014, BremerWeberTierny2010} that have potential use in TC tracking. 
Most approaches identify features of interest with regions enclosed by streamlines or level sets (\ie, isosurfaces or contours).
Wischgol~\etal~\cite{WischgollScheuermann2001} discussed closed streamlines in 2D fields, which can represent the eyes of the TCs and indicate locations and sizes of the eyes in TC tracking. 
Correspondences between features in consecutive time steps can be identified via spatial overlap~\cite{SohnBajaj2005}, critical point tracking~\cite{SkrabaWang2014}, tree structure~\cite{BremerWeberTierny2010}, or combinatorial feature flow fields~\cite{ReininghausKastenWeinkauf2012}. Critical point tracking is most relevant to our framework, which is reviewed next. 
}

\subsection{Vector Field Topology: Critical Point Tracking}
Critical point tracking establishes the correspondences between critical points in successive time steps. 
It is a key tool from vector field topology and plays an important role in understanding the behavior of time-varying vector fields.
Algorithms for critical point tracking may be classified as proximity-, integral-, and interpolation-based methods. 
Proximity-based methods (e.g.,~\cite{HelmanHesselink1989, HelmanHesselink1990}) find correspondences of critical points based on distance proximity in the domain. 
Integral-based approaches represent the tracking of critical points as streamlines of a higher dimensional field, called the feature flow field (FFF)~\cite{TheiselSeidel2003,WeinkaufTheiselVan-Gelder2010, ReininghausKastenWeinkauf2012}, and compute feature tracks based on tangent curves in FFF. 
Interpolation-based methods take into account the time as an additional dimension in addition to the space domain~\cite{TricocheScheuermannHagen2001a,TricocheWischgollScheuermann2002, GarthTricocheScheuermann2004,GuoLenzXu2021}.

The robustness of critical points has been introduced to quantify the structural stability of critical points~\cite{SkrabaRosenWang2016} and has been used in vector field simplification~\cite{SkrabaWangChen2014, SkrabaWangChen2015}, feature extraction~\cite{WangBujackRosen2017}, and visualization~\cite{WangRosenSkraba2013}. 
Skraba and Wang~\cite{SkrabaWang2014} showed the potential use of robustness in feature tracking, that is, finding correspondences between critical points based on their closeness in stability, measured by robustness, instead of just distance proximity within the domain.
Building on the theoretical basis established by~\cite{SkrabaWang2014}, Yan~\etal~\cite{YanUllrichVan-Roekel2022} proposed a multilevel robustness ({\MR}) framework to realize critical point tracking in practice for large-scale scientific simulations, see~\cref{sec:ml-Robustness} for details.

\section{Technical Background}
\label{sec:background}

We first review the classic notion of robustness and multilevel robustness, which are customized to build {\tool} (see \cref{sec:TROPHY}).

\subsection{Robustness}
\label{sec:classicRobustness}

\para{Critical points of a 2D vector field.} 
Unless otherwise specified, we work with a 2D \emph{vector field} $f: \Xspace \subseteq \Rspace^2 \to \Rspace^2$, which assigns a 2D vector to each point in $\Xspace$. 
We use $u_{10}$ and $v_{10}$ to represent the 10-meter zonal (west-east) and meridional (south-north) wind vector components, respectively.  
Then, $f$ is expressed as $f(x) = (u_{10}(x), v_{10}(x))^T$. 

\begin{figure}[t]
    \centering
    \includegraphics[width=\columnwidth]{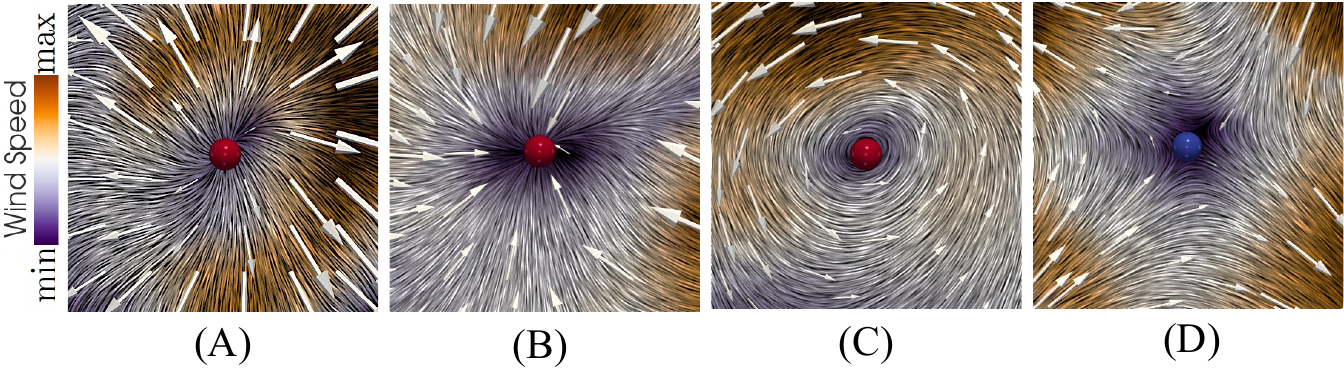}
    \vspace{-6mm}
\caption{Four types of 2D critical points: (A) source, (B) sink, (C) center, and (D) saddle. Each point is colored by its degree: red means degree $+1$, blue means degree $-1$. Vector fields are visualized by the line integral convolution (LIC)~\cite{CabralLeedom1993}, colored by wind speed, with sampled arrow glyphs (the length of an arrow is proportional to the wind speed). }
    \label{fig:CriticalPoints}
    \vspace{-4mm}
\end{figure}

A \emph{critical point} $x \in \Xspace$ in $f$ is where the vector vanishes, that is, $|f(x)| = 0$. 
A critical point $x$ can be classified {\wrt} its \emph{degree} $\mydeg(x)$, defined as the number of field rotations while traveling along a closed curve counterclockwise surrounding $x$ (enclosing no other critical point). 
A source/sink/center has degree $+1$, whereas a saddle point has degree $-1$.
Critical points are important features in studying flow behavior in many applications; see \cref{fig:CriticalPoints} as an example.
%\cref{fig:CriticalPoints} shows four types of critical points from the $\EW$ dataset (see~\cref{sec:data} for details). 

\begin{figure*}[t]
    \centering
    \includegraphics[width=2.01\columnwidth]{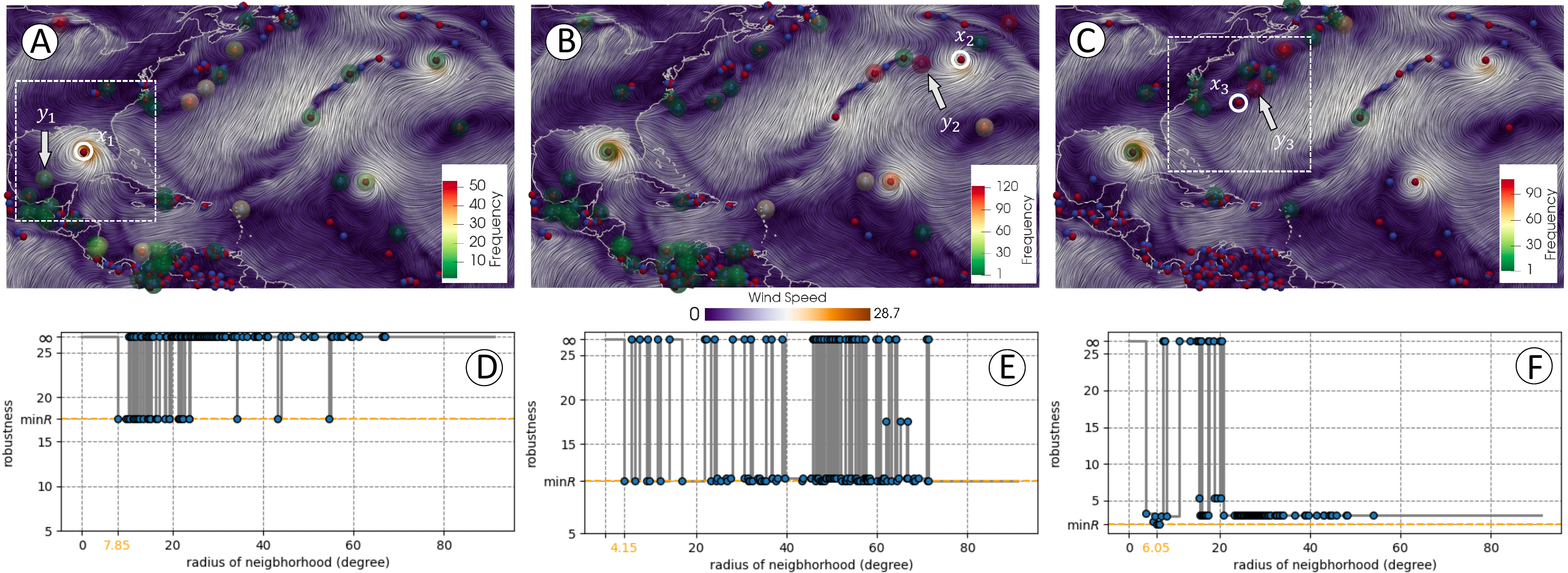}
    \vspace{-2mm}
\caption{Cancellation partners and the exact multilevel robustness of $x_1$, $x_2$, and $x_3$. (A-C) Cancellation partners of a selected critical point are wrapped in bubbles and colored by their frequencies as partners. (D-F) The exact multilevel robustness of $x_1$, $x_2$, and $x_3$.}
    \label{fig:Partners}
     \vspace{-2mm}
\end{figure*}

In most cases, the center of a TC can be detected as a center in a vector field when it is intensified into a strong hurricane, with very low wind speed in the eye and extremely high wind speed along the eyewall. 
During the dissipating phase of a TC, such as at its landfall, the center of the TC can be detected as either a source or a sink. 
If a center transforms to a source, then it indicates a divergence in meteorology, which means the weather can be clear and calm. 
If a center transforms to a sink, then it indicates a convergence, which is associated with clouds and precipitation. 
An example of this phenomenon is Hurricane Florence in 2018: a clear sink forms in the 2D wind field  during its landfall, bringing 1-in-500-year expected flooding due to heavy precipitation.

%and can be detected as source points when it is weaken on its dissipating phase, such as at its landfall, \jiali{sentences after this need to be changed if there is no sink. what i wrote there is not true for source, it's true for sink..} which can still bring severe weather such as heavy precipitation, as the source is still a low pressure system, favorable for air rising, cooling and condensation.
 
% Arrow glyphs are placed on sampled points in the domain to indicate the directions of vectors. 

\para{Merge trees.} 
The computation of robustness relies on an \emph{augmented merge tree} modified from the classic merge tree. 
Given a scalar function $f_0$ defined in a 2D domain $\Xspace$, $f_0:\Xspace \to \Rspace$, let $\Xspace_r=f_0^{-1} (-\infty, r]$ denote the \emph{sublevel set} of $f_0$ for some $r \geq 0$. 
A classic \emph{merge tree} is constructed by tracking the evolution of (connected) components in $\Xspace_r$ as $r$ increases. Leaves in a merge tree represent the creation of a component at a local minimum of $f_0$, internal nodes represent the merging of components, and the root represents the entire space as a single component.

\begin{figure*}[t]
    \centering
    \includegraphics[width=2.01\columnwidth]{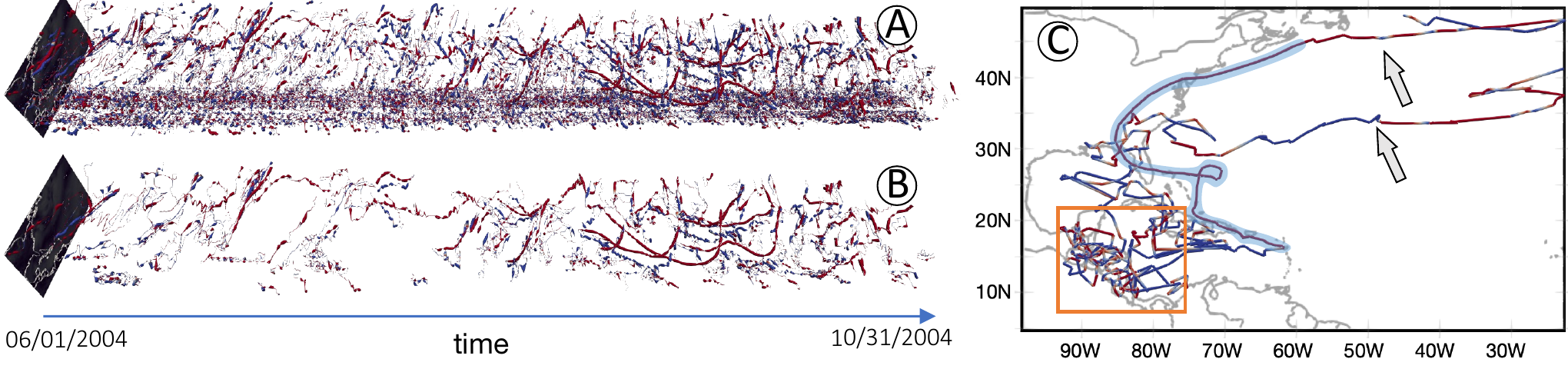}
    \vspace{-4mm}
\caption{FTK tracking result of the $\EFour$ dataset.
(A) The original FTK tracking result. (B) The filtered FTK tracking result after physics-informed feature selection. (C) One track from (A). The radius of a track in (A) and (B) is proportional to its classic robustness.  Sources/sinks/centers are in red, and saddles are in blue.}
    \label{fig:filterBeforeMR}
    \vspace{-2mm}
\end{figure*}

\begin{figure*}[t]
    \centering
    \includegraphics[width=2.01\columnwidth]{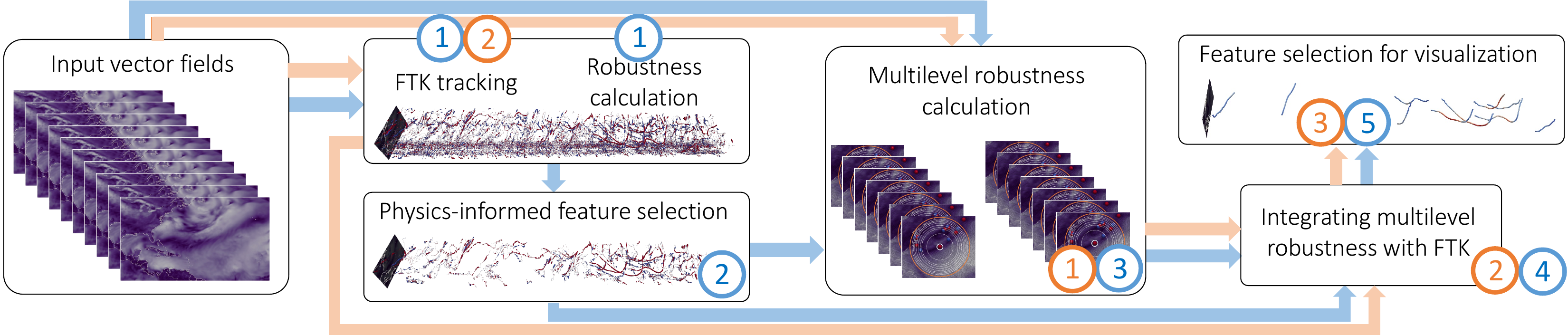}
\caption{Implementation pipelines for the {\MR} (in orange) and {\tool} (in blue). Numbers 1-5 correspond to the steps for each framework.}
    \label{fig:pipeline}
    \vspace{-4mm}
\end{figure*}

\myedit{To construct an augmented merge tree from a 2D vector field $f$, first we define a scalar field $f_0:\Xspace \to \Rspace$ by assigning the vector magnitude to each point $x \in \Xspace$, that is, $f_0(x) = ||f(x)||_2$. In this paper, $f_0$ can be expressed as wind speed.
Second, instead of using local minima of $f_0$ as leaves of the merge tree, the leaves of our augmented merge tree consist of $\Xspace_0$, which is precisely the set of critical points of $f$. 
The tracking of the merging behavior of components is the same as classic merge tree construction. 
Third, once the merge tree is constructed, it can be further augmented with the degrees of critical points (on leaves) and the degrees of components (on internal nodes). The degree of a component is defined as the sum of degrees of critical points the component contains. See~\cref{sec:exampleMT} for an example.}

\para{Robustness calculation.}
The topological notion of \emph{robustness} quantifies the stability of a critical point \wrt~perturbations of the vector field. Let us define the concept of vector field \emph{perturbation} first. A continuous mapping $h: \Xspace \to \Rspace^2$ is an \emph{$r$-perturbation} of $f$, if $d(f, h) \leq r$, where $d(f, h)=\sup_{x\in \Xspace}||f(x)-h(x)||_2$, and $\sup$ means supremum. 
See~\cite{WangRosenSkraba2013} for some mathematical properties of robustness and lemmas to support critical points cancellation under vector field perturbation. 

The robustness of a critical point can be calculated as the function value of its lowest zero-degree ancestor in the augmented merge tree~\cite{WangRosenSkraba2013}. \myedit{See~\cref{sec:exampleRob} for an example.}

%which ignores the possibility of the occurrences of perturbation within a local neighborhood.
%; see~\cite[Sect. 4]{YanUllrichVan-Roekel2022} for an example. 

\subsection{Multilevel Robustness}
\label{sec:ml-Robustness}
In practice, vector fields generated from large-scale ocean and atmospheric datasets contain features at different scales. The drawback of classic robustness comes from building a single merge tree with critical points in the entire domain, which suffer from undesirable boundary effects~\cite{YanUllrichVan-Roekel2022}. 
To mitigate such drawbacks, Yan~\etal\cite{YanUllrichVan-Roekel2022} introduced a notion of multilevel robustness (reviewed in~\cref{sec:MRdef}).
This notion captures the multiscale nature of the data and mitigates the boundary effects suffered by classic robustness computation. 
It also shows initial promise in critical point tracking in practice. 
%They also proposed a multilevel robustness framework to realize the robustness-based critical point tracking in practice. 
We review how the notion of multilevel robustness can improve the feature-tracking results in~\cref{sec:intrgrateWithFTK}, and we give the pipeline to implement the multilevel robustness framework in~\cref{sec:pipelineMR}. For simplicity and comparative purposes, we refer to this original multilevel robustness-based tracking~\cite{YanUllrichVan-Roekel2022} as the {\MR} framework in the remainder of this paper.

\subsubsection{The Multilevel Robustness}
\label{sec:MRdef}
Roughly speaking, the multilevel robustness of a critical point $x \in \Xspace$ can be defined as a sequence of robustness values computed from its neighborhoods of increasing radii. 
Formally, let $B_x(a)$ denote a ball of radius $a$ with a critical point $x \in \Xspace$ as its center.
The multilevel robustness of $x$ can be expressed as 
$R_x: [0,\infty) \to \Rspace$, where $R_x(a)$ is the (classic) robustness of $x$ computed {\wrt} the domain $B_x(a)$ for $a \in [0,\infty)$.
%see~\cref{fig:AdaptiveRegions} for multiple neighborhoods of $x_1$ and $x_3$ with different radii.
Assuming the domain $\Xspace$ contains $n$ critical points, then for a fixed critical point $x \in \Xspace$, its multilevel robustness will change at most $n-1$ times as $a$ increases, since $x$ gets one more candidate as its the cancellation partner as $B_x(a)$ passes through each critical point. 

In~\cref{fig:Partners} (D-F), we give the exact multilevel robustness of $x_1$, $x_2$, and $x_3$, respectively, where the $x$-axis corresponds to the increasing radii and the $y$-axis represents their classic robustness values. We highlight the radii when the neighborhood includes new critical points with blue points in~\cref{fig:Partners} (D-F).  
In~\cref{fig:Partners} (A-C), we visualize all cancellation partners for selected critical points when we use different sizes of neighborhoods in classic robustness computation. 
The cancellation partners are wrapped in bubbles and colored by the number of times that are referred to as cancellation partners of selected critical points. 
For example, $x_1$ and $y_1$ are paired as cancellation partners 12 times, whereas $x_2$ and $y_2$ are paired 122 times.
The classic robustness of $x_1$ calculated with the entire  domain is infinity, even if it can be canceled with $y_1$ within a $7.85$-degree region under a $17.6$-perturbation. 
This phenomenon happens because $x_1$ represents the center of a large-scale cyclone and is surrounded by flows of a large magnitude. If we build an augmented merge tree from the entire input domain during classic robustness calculation, the lowest ancestor of $x_1$ will be the ancestor of the most critical points in the domain. 
This limitation explains why $x_1$ has potential cancellation partners across the entire domain and may not be able to find its cancellation partner if the degree of the entire domain is not equal to zero. See~\cite[Fig. 2]{YanUllrichVan-Roekel2022} for another example. 
 
Therefore, the drawback of classic robustness comes from building a single merge tree with critical points in the entire domain, which ignores the possibility of the occurrences of cancellation within a local neighborhood. The definition of multilevel robustness successfully captures the multiscale nature of the data and mitigates the drawbacks of the classic robustness computation. However, computing the multilevel robustness exactly is time-consuming. For the vector field containing $n$ critical points, we need to conduct $n \times (n-1)$ classic robustness computations. 
In~\cite{YanUllrichVan-Roekel2022}, the {\MR} framework approximates the exact multilevel robustness by using $N$-level robustness. That is, for a critical point $x \in \Xspace$, the authors considered $N$ number of its neighborhoods at radius $\{a_0, \dots, a_{N-1}\}$, where each $a_i := L \times (i+1)/N$ and $L$ is the diameter of the domain $\Xspace$. 
In this case, the approximations of multilevel robustness for all critical points require $n \times N$ classic robustness computations and work well in their applications.

\subsubsection{Enhancing Feature Tracking with Multilevel Robustness}
\label{sec:intrgrateWithFTK} 
The multilevel robustness can be integrated with any existing feature-tracking algorithms to improve the understanding of vector field dynamics.
Yan et al.~\cite{YanUllrichVan-Roekel2022} utilized the minimum multilevel robustness $\minR_{x} := \min_{a \in [0,~L)} R_x(a) $ for their visualization tasks, since $\minR_{x}$ approximates the smallest possible amount of perturbation to the vector field necessary to cancel each critical point. 
The authors 
%\cite{YanUllrichVan-Roekel2022} 
integrated the $\minR_{x}$ with FTK~\cite{GuoLenzXu2021}, a state-of-the-art feature-tracking technique. 
We also utilize FTK in {\tool}.

The initial critical point tracks from FTK suffer from visual clutter when we deal with large-scale datasets. 
\cref{fig:filterBeforeMR} (A) shows the FTK tracking result for the $\EFour$ dataset whose time steps range from 06/01/2004 to 10/31/2004 with a six-hour time gap (see~\cref{sec:evaluation,sec:data} for details). 
Because of visual clutter among thousands of tracks, it is hard for us to identify the dominant features. 
Since the FTK algorithm considers only the correspondences of critical points based on 0-levelset extraction, some important features (\eg,~centers of cyclones) will be included in the same track with other noisy features. 
\cref{fig:filterBeforeMR} (C) shows one of the FTK tracks from \cref{fig:filterBeforeMR} (A). 
This long track contains a Category 3 hurricane, named Jeanne, as highlighted with the blue curve in~\cref{fig:filterBeforeMR} (C). However, it also contains unstable features on the Gulf of Mexico; indicated within the orange box of ~\cref{fig:filterBeforeMR} (C). 

The main idea of enhancing feature tracking with multilevel robustness is to segment and reconnect the initial tracks obtained by FTK considering the minimum multilevel robustness. 
The {\MR} framework can break initial FTK tracks into more meaningful segments with similar robustness values.  
In the example of~\cref{fig:filterBeforeMR} (C), the {\MR} framework can extract the part highlighted with the blue curve from the other part of the track.
This framework can also remove unstable features in the middle of a meaningful track and reconnect remaining parts as a new track after examining spatial faces and spacetime edges~\cite{GuoLenzXu2021} of breakpoints; see~\cite[Sect. 5.1]{YanUllrichVan-Roekel2022} for a concrete example.

\subsubsection{Pipeline of the Multilevel Robustness Framework}
\label{sec:pipelineMR}
As shown in~\cref{fig:pipeline} (orange arrows and indices), the implementation of {\MR} framework involves the following three steps: 

\para{Step 1:~multilevel robustness calculation.} 
%Multilevel robustness is defined as a sequence of robustness values computed from its neighborhoods with increasing radii. 
The {\MR} framework calculates the multilevel robustness for all detected critical points with evenly increased radii until the neighborhood includes the entire input domain. 
Then, the minimum multilevel robustness is calculation for postprocessing.

\para{Step 2:~integration with feature tracking.}
The {\MR} framework integrates the minimum multilevel robustness with FTK~\cite{GuoLenzXu2021} to enhance the original FTK tracking results.
%The main idea is to add a step to segment/reconnect the original FTK tracks into more meaningful segments, based on the multilevel robustness.

\para{Step 3:~feature selection.}
The {\MR} framework utilizes two filters based on multilevel robustness and degree information of tracks for feature selection. 
These feature selection strategies can help users reduce visual clutter and highlight dominant features in the domain.

{\tool} 
%inherits two key capabilities of the multilevel robustness framework: capture the multiscale nature of the data and enhance critical points tracking results to understand the vector field dynamics. It 
reuses the notion of multilevel robustness, described in~\cref{sec:MRdef}, and the method to integrate the minimum multilevel robustness with FTK; see~\cref{sec:intrgrateWithFTK}. 
In the following section, we customize the {\MR} framework to {\tool} by integrating the physical knowledge of TCs in feature extraction and tracking.

\section{Method: {\tool} for Cyclone Tracking}
\label{sec:TROPHY}

Our physics-informed tracking framework, {\tool}, encodes several criteria considering the physical properties of cyclones.  
These criteria make {\tool} more efficient and appropriate for cyclone tracking than the {\MR} framework. 

\para{Overview of {\tool}.}
An overview of our pipeline is shown in~\cref{fig:pipeline}. 
First, we compute the FTK tracking result for the whole dataset and the classic robustness of the critical points using the entire domain for each time step. 
Second, the FTK tracks and the robustness of critical points are used in the physics-informed feature selection (\cref{sec:FeatureSeclectionBeforeMR}) to filter out noise-liked tracks. 
Third, an adaptive-level strategy (\cref{sec:adaptiveLevel}) is applied in the multilevel robustness calculation for selected critical points.
Fourth, {\tool} integrates the minimum multilevel robustness with FTK to enhance the original FTK tracking results. 
This step uses the same strategy with the {\MR} framework; see~\cref{sec:intrgrateWithFTK} or~\cite[Sect. 5.1]{YanUllrichVan-Roekel2022} for a detailed discussion.
Finally, we utilize one \emph{stability filter} function from~\cite{YanUllrichVan-Roekel2022} and propose two additional physics-informed filter functions to highlight cyclones; see~\cref{sec:FeatureSelection}.

\subsection{Physics-Informed Feature Selection}
\label{sec:FeatureSeclectionBeforeMR}

We now introduce a physics-informed feature selection strategy during preprocessing, making {\tool} much more efficient than the {\MR} framework in studying large-scale datasets.  
The {\MR} framework computes multilevel robustness for all detected critical points. 
For the example in~\cref{fig:filterBeforeMR} (A), 102,720 critical points and 22,743 tracks are detected with FTK. 
Suppose we set the number of levels $N= 50$, the {\MR} framework needs to conduct $102720 \times 50$ times classic robustness computation. 
{\tool} instead focuses on tracking cyclone-liked features, which selects a subset of critical points/tracks for in-processing. 
Considering the physical properties of real-world cyclones, we incorporate two criteria in physics-informed feature selection before multilevel robustness calculations. 

First, since the duration of a tropical storm/cyclone is usually larger than 1 day, we require the selected tracks to contain at least one 1-day segment that consists of $+1$ degree critical points only. 

Second, since a critical point representing the center of a cyclone usually has a high stability measure across time before it hits the land and dissipates, we require the segment detected from the previous step to have a high average robustness value.
Based on domain knowledge, a tropical storm must have maximum sustained winds of at least 17.5 m/s. 
We set the threshold to 1.75 for track filtering. 

With these two requirements, the numbers of critical points and tracks decrease to 9,784 and 86, respectively; compare~\cref{fig:filterBeforeMR} (A) and (B). 
Thus, {\tool} needs to compute multilevel robustness for only $9.5\%$ of critical points compared with the original {\MR} framework.

\begin{figure}[t]
    \centering
    \includegraphics[width=\columnwidth]{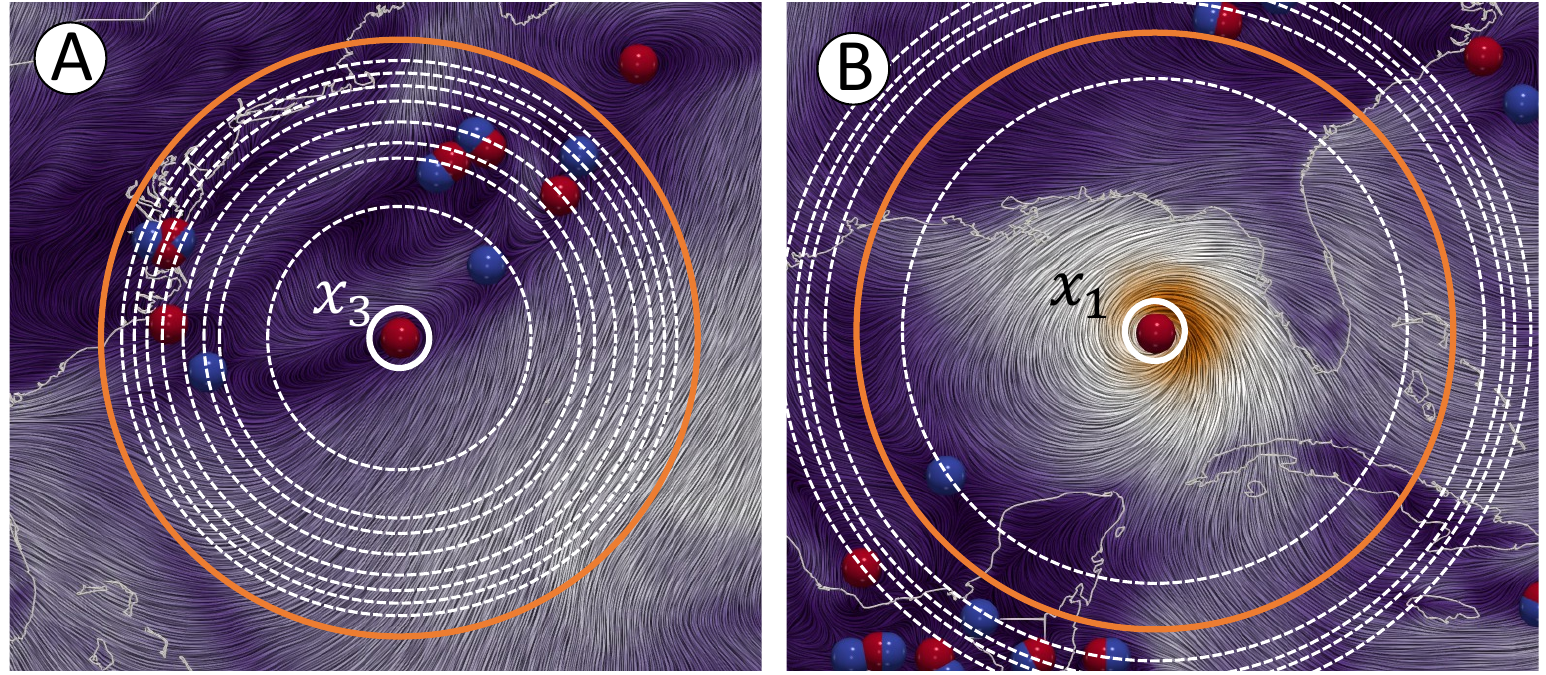}
    \vspace{-4mm}
\caption{Adaptive levels of neighborhoods for critical points $x_3$ and $x_1$. (A) and (B) are truncated regions from~\cref{fig:Partners} (A) and (C).}
    \label{fig:AdaptiveRegions}
    \vspace{-4mm}
\end{figure}

\subsection{Adaptive Levels for Multilevel Robustness}
\label{sec:adaptiveLevel}

We next propose an adaptive level strategy for the multilevel robustness calculation. 
Such a strategy also considers physical properties of real-world cyclones and leads to more reasonable minimum multilevel robustness values for {\tool} than does the {\MR} framework. 
As discussed in~\cref{sec:MRdef}, the {\MR} framework approximates the exact multilevel robustness with a set of evenly spaced radii. 
This strategy may lead to two consequences that are counterintuitive in real-world cyclone analysis. 
First, a critical point can be canceled with other critical points that are far away in the known data domain due to boundary effects; see partners for $x_1$ and $x_2$ in~\cref{fig:Partners} (A) and (B). 
Although such cancellations are mathematically justifiable, it is almost impossible to happen in real-world scenarios   since no perturbation could happen across the whole Atlantic Ocean. 
Second, the {\MR} framework may not find the true minimum multilevel robustness value due to sampling. 
It may also waste computational resources when an enlarged neighborhood does not include potential cancellation partners. For example, increasing the radius of the neighborhood from 60 to 80 for $x_3$ in~\cref{fig:Partners} (C) does not lead to new cancellation candidates.  

To mitigate the drawbacks of the {\MR} framework, we introduce an adaptive-level strategy. 
First, we set a physics-informed neighborhood size, where real-world perturbation could happen. 
We set this radius to be 10 degrees since hurricanes are among the most destructive real-world perturbations to wind field and are typically about 4.7 degrees wide. 
The {\tool} considers all possible cancellation partners within this neighborhood using varying radii.  
\cref{fig:AdaptiveRegions} (A) illustrates our adaptive level strategy in calculating the multilevel robustness. 
For a critical point $x$, we consider all critical points within its neighborhood at radius 10. 
Suppose there are $N$ ($N \ge 10$) critical points in this neighborhood and their Euclidean distances to $x$ are $\{a_0, \dots, a_{N-1}\}$. We compute the classic robustness of $x$ within neighborhood defined by these radii, giving rise to its multilevel robustness. 
If $N < 10$, we select an additional $10-N$ closest critical points outside of the selected neighborhood for more candidates of cancellation partners; see~\cref{fig:AdaptiveRegions} (B). 
However, in most cases, the cancellation partners for the true minimum multilevel robustness are located in our physics-informed neighborhood, for example, $7.85$ degree for $x_1$, $4.15$ degree for $x_2$, and $6.05$ degree for $x_3$ in~\cref{fig:Partners}.

\begin{figure}[b]
    \centering
    \includegraphics[width=\columnwidth]{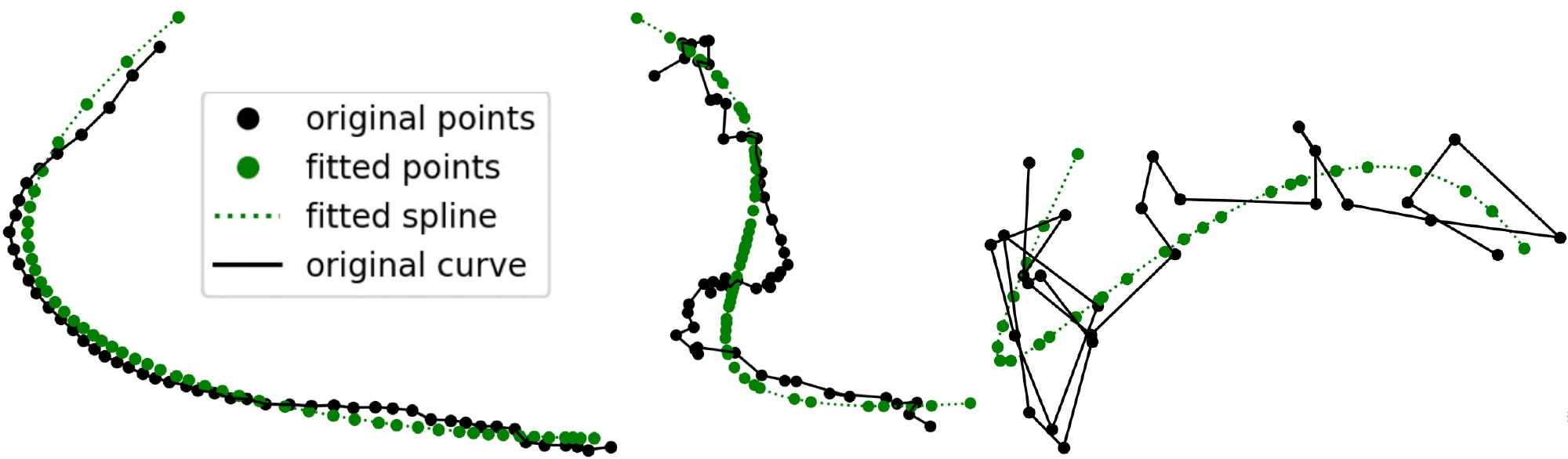}
    \vspace{-4mm}
\caption{Original trajectories and their fitted splines.}    \label{fig:FittingCurve}
    \vspace{-4mm}
\end{figure}

\subsection{TC Feature Selection for Visualization} 
\label{sec:FeatureSelection}

We now present feature selection aided by the minimum multilevel robustness and physical properties of TCs.
We inherit one \emph{stability filter} from the {\MR} framework.  
This filter considers the minimum multilevel robustness $\minR_x$ and its temporal stability in terms of lifespan. 
Let $l$ denote a logistic transformation of $\minR_x$, which maps the $\minR_x \in [0, \infty]$ to $l(\minR_x) \in [0,1]$. This normalization is defined as 
 \[
 l(\minR_x)=\frac{2}{1+e^{-k\cdot \minR_x}}-1,
 \]
where $k$ is the logistic growth rate; see~\cite[Sect. 5.1]{YanUllrichVan-Roekel2022} for a detailed discussion on the benefit of this normalization and parameter selection for $k$.
Now let $\gamma$ denote a track,  $|\gamma|$ is its total length. 
The \emph{stability} of a track $\gamma$ is defined as
\begin{equation}
\label{eqn:stability}
  b(\gamma) := \frac{\sum_{x \in \gamma}l(\minR_x)}{|\gamma|} \cdot \frac{t_\gamma}{T},
\end{equation}
where $T$ is the temporal span of the input dataset and $t_\gamma$ is the temporal span of $\gamma$.
The first term in Eq.~\eqref{eqn:stability} captures the average pointwise stability (in a logistic scale), whereas the second term encodes the lifespan of the track. 
By definition, $b(\gamma) \in [0, 1]$. 
 
Our second feature selection strategy is referred to as  \emph{maximum wind speed} (MWS) \emph{filter}.  
Since the storm systems are usually categorized by the Saffir--Simpson Hurricane Wind Scale, which considers only a hurricane's maximum sustained wind speed, we use this filter to control the categories of storm that users want to visualize.
Formally, for a track $\gamma$, its \emph{maximum wind speed} is
\begin{equation}
\label{eqn:windSpeed}
  w(\gamma) := \max_{x \in \gamma} \omega(x),
\end{equation}
where $\omega(x)$ is the MWS within the two-degree region of the center $x$. 

Our third feature selection strategy is referred to as  \emph{smoothness filter}. 
It is used to filter out tracks that are too tortuous to be considered as TC tracks. 
Roughly speaking, we define the smoothness of a track by the average distance between the normalized track and its smooth univariate spline. 
Formally, for a track $\gamma$, its \emph{smoothness} is
\begin{equation}
\label{eqn:smooth}
  s(\gamma) := 1-\frac{\sum_{x \in \gamma}||J(x)-U(J(x))||}{|\gamma|} ,
\end{equation}
where $J$ is a normalization term mapping $x$ to $[0,1]\times [0,1]$ and $U$ maps $J(x)$ to the point on the smooth univariate spline of the normalized $\gamma$. 
{\tool} uses the SciPy Python library to calculate the (degree 3) smooth univariate splines for detected tracks.
\cref{fig:FittingCurve} shows three curves from our climate dataset and their smooth univariate splines. The left track has the lowest average distance and hence the highest smoothness value, whereas the right track has the lowest smoothness value and usually cannot be considered as a TC track.

\begin{figure*}[t]
    \centering
    \includegraphics[width=2.01\columnwidth]{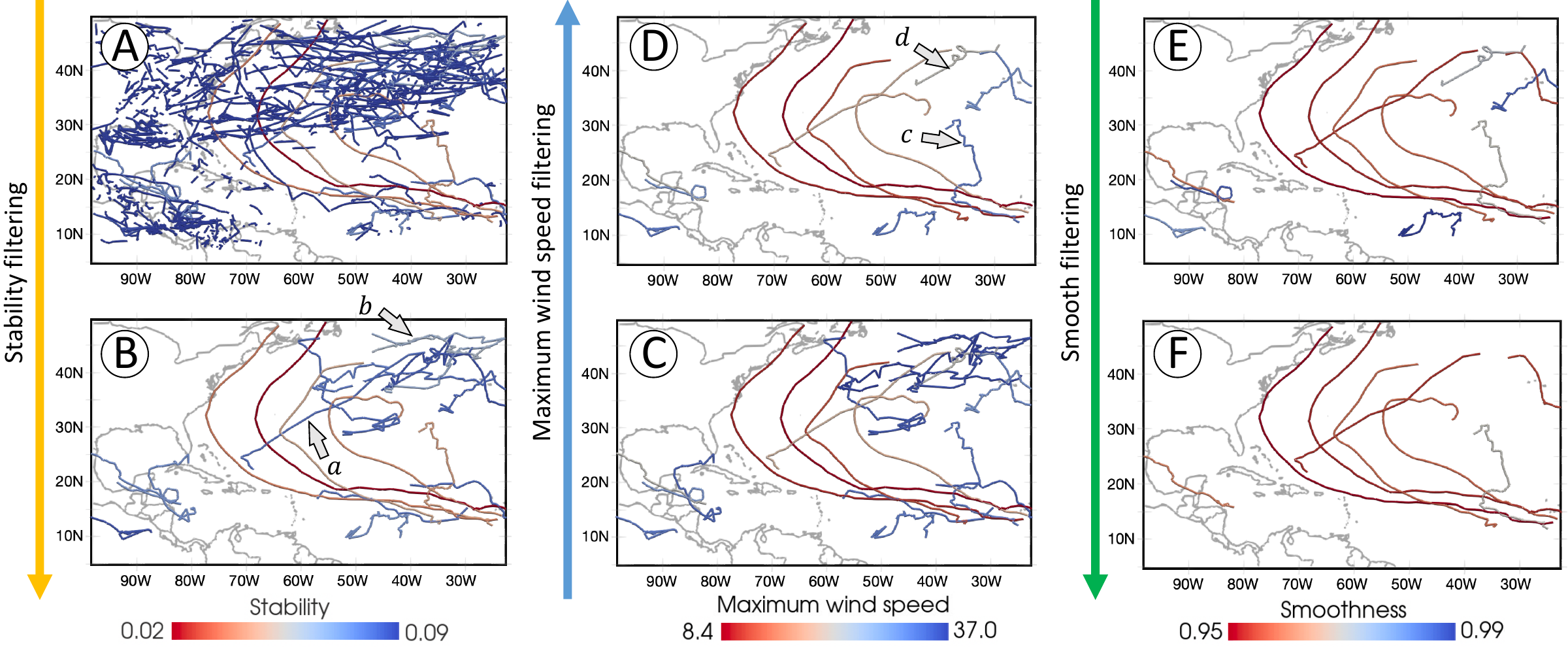}
    \vspace{-2mm}
\caption{Feature selection for the $\ETen$ dataset. Tracks are filtered and colored by their stability (A$\rightarrow$B), maximum wind speed (C$\rightarrow$D), and smoothness (E$\rightarrow$F).}
    \label{fig:filters}
    \vspace{-4mm}
\end{figure*} 

In~\cref{fig:filters}, we apply the above three filters to the $\ETen$ dataset (see~\cref{sec:evaluation,sec:data} for details). After integrating multilevel robustness with FTK tracking results, we obtain tracks in~\cref{fig:filters} (A), which still suffer from visual clutter.
By applying the stability filter with a threshold of $0.029$ for $b(\gamma)$, we obtain tracks in~\cref{fig:filters} (B). 
If we keep enlarging the stability threshold, track $a$ will be filtered out before track $b$, even if track $a$ represents the Category 1 hurricane Otto, and track $c$ cannot be found in the National Hurricane Center’s Tropical Cyclone Reports. 
Therefore, we use our MSF filter $w(\gamma)$ based on the MSF along the track and set the threshold to be $13.5$. 
The remaining tracks are shown in~\cref{fig:filters} (D). 
Again, if we keep increasing the threshold for $w(\gamma)$, track $c$ will be filtered out before track $d$, whereas track $c$ represents the Category 1 hurricane Lisa. 
Therefore, we apply our third smoothness filter $s(\gamma)$ in~\cref{fig:filters} (E) and (F). 
Now, {\tool} highlights 8 tracks representing hurricanes/tropical storms that can all be found in the National Hurricane Center’s Tropical Cyclone Reports.

\section{Metrics, Data, and Methods for Evaluation}
\label{sec:evaluation}

We present experimental results using using 30-year (1981--2010) near-surface wind vector field from the ECMWF Reanalysis v5 (ERA5). We annotate this 30-year dataset as the $\EW$ dataset. We also mark the one-year subset data from the $\EW$ dataset as $\EY$.
We use the International Best Track Archive for Climate Stewardship (IBTrACS~\cite{KnappKrukLevinson2010} version 4) observations as the reference and the TC tracking results of the \TE~software package~\cite{UllrichZarzycki2017} for comparision. 
The details of datasets, IBTrACS, and \TE~are described in~\cref{sec:data}.
We now describe the metrics used for evaluating the performance of {\tool} in~\cref{sec:metrics}. 
We also discuss the parameter tuning for {\tool} in~\cref{sec:parameter}.

\subsection{Metrics for Evaluating Tropical Cyclones}
\label{sec:metrics}

We review several metrics used for evaluating TCs in climate data. See~\cite{ZarzyckiUllrichReed2021} for detailed descriptions. 

\begin{figure*}[t]
  \centering
  \includegraphics[width=\linewidth]{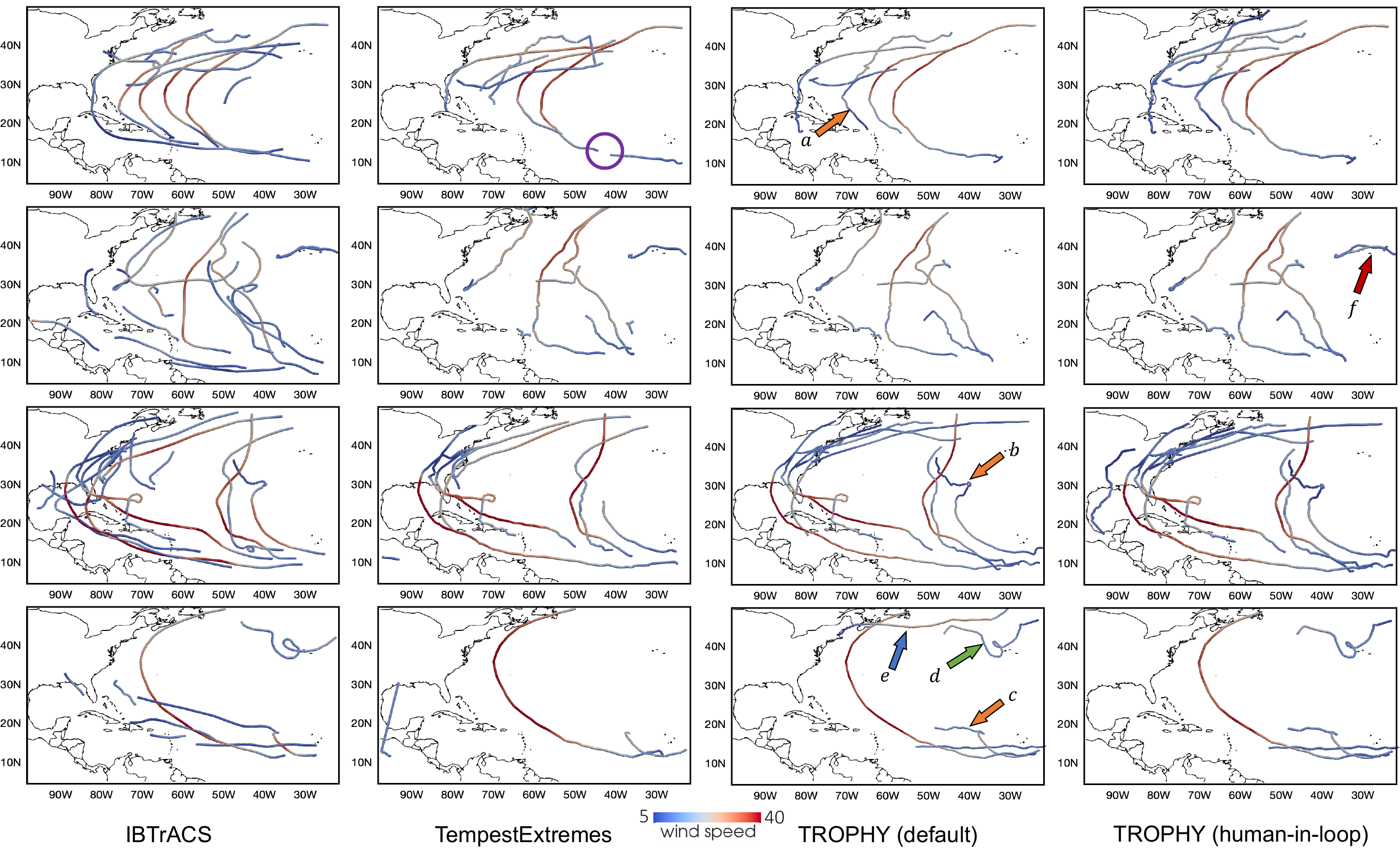}
   \vspace{-4mm}
  \caption{
   Annually detected TCs by IBTrACS (1st column), \TE (2nd column), the {\tool} (3rd column) with default setting, and {\tool} (4th column) with human-in-the-loop in 1981 (1st row), 1990 (2nd row), 2004 (3rd row), and 2009 (4th row). Tracks are colored by maximum wind speed within the two-degree great circle of the detected TC eyes. 
  }
   \vspace{-4mm}
  \label{fig:teaser}
\end{figure*}

\subsubsection{Storm Climatology and Characteristics}
\para{Annual frequency}, marked as count or $m$ (\#), is measured by the number of discrete storm events.

\para{Annual duration}, marked as $\TCD$ (days), can be defined as 
$ \TCD_m = \frac{1}{4}\sum_{i\in [1, m]} \ocr_{6h, i}$,
where $\ocr_{6h, i}$ is occurrence of 6 hourly-tracked points during the lifetime of storm $i$. 

\para{Storm genesis}, marked as gen, is defined as the first entry for each individual storm’s lifetime.

\para{Storm intensity} can be measured by the minimum sea level pressure at the cyclone center, marked as SLP (hPa), and two-degree maximum 10-meter wind speed, marked as $u_{10}$ ($m/s$).

\para{Latitude of lifetime-maximum intensity}, marked as LMI, is defined as the absolute value of the latitude where a TC reaches its maximum intensity (as defined by maximum $u_{10}$). 

\subsubsection{Statistics}
We employ two statistical techniques to evaluate the above metrics on a 30-year (1981--2010) dataset.

One is the \emph{arithmetic mean}, $\bar{x} = \frac{1}{n}\sum_{i=1}^n x_i$,
which is used in~\cref{sec:AnnualClimatology} to study annual domain-averaged climatology. 
%For example, $\overline{\TCD}$ represents the average annual duration of storms.

The second is the \emph{Pearson correlation} coefficient $r_{xy}$,
 defined as
%\begin{equation}
%\label{eq:PearsonCor}
\[
 r_{xy} = \frac{\sum_{i=1}^n (x_i-\bar{x})(y_i-\bar{y})}{\sqrt{\sum_{i=1}^n (x_i-\bar{x})^2}\sqrt{\sum_{i=1}^n (y_i-\bar{y})^2}},
 \]
%\end{equation}
%where $x$ can be measured items from {\tool} and $y$ can be items from reference. 
which is used in~\cref{sec:spatialClimatology} to evaluate the similarity of metrics {\wrt} temporal (\eg, storm frequency) or spatial (\eg, genesis density) patterns, which are generated by TC tracking results from the tested tracking algorithm and reference observations.

\subsection{Configuration for Feature Selection}
\label{sec:parameter}
For filters introduced in~\cref{sec:FeatureSelection}, we recommend a default value as a threshold for each filter function based on the TC tracks provided by IBTrACS.
First, we calculate the spatial pattern of cumulative track density using TC tracks detected by {\tool} with thresholds of stability $b(\gamma)$ varying from 0.02 to 0.035, MWS $w(\gamma)$ varying from 10 to 14, and smoothness $s(\gamma)$ varying from 0.95 to 0.97.
We also calculate the spatial pattern of cumulative track density from IBTrACS as a baseline. 
Then, we evaluate the similarity between the density patterns from {\tool} and IBTrACS using Pearson correlation, marked as $r_{xy,track}$, which is the most exhaustive measure of TC activity.
We provide a detailed example for $r_{xy,track}$ computation in~\cref{sec:spatialClimatology}.
We suggest the thresholds for $b(\gamma)$, $w(\gamma)$ , and $s(\gamma)$ with the values when $r_{xy,track}$ reaches its maximum, that is, 0.029 for stability $b(\gamma)$, 13.5 for MWS $w(\gamma)$, and 0.967 for smoothness $s(\gamma)$. 
{\tool} also allows users to fine-tune these thresholds independently for individual TC tracks of interest. 
We report performances of {\tool} with both default thresholds and human-in-the-loop thresholds in~\cref{sec:results}. 
We demonstrate that, although the tracking results using default threshold are encouraging, results with the human-in-the-loop option can further improve the performance of {\tool}. 
\section{Cyclone Tracking Results With {\tool}} 
\label{sec:results}
We demonstrate cyclone-tracking results using {\tool} and their comparisons with a state-of-the-art TC tracking algorithm. Observations from official forecast centers are provided for reference.

\begin{figure*}[t]
    \centering
    \includegraphics[width=2.01\columnwidth]{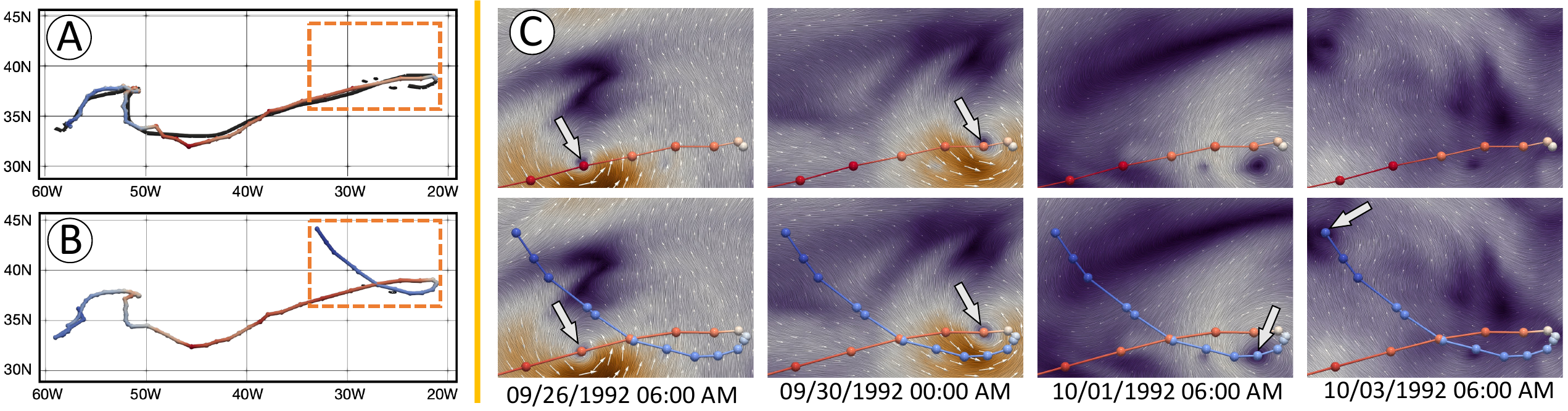}
    \vspace{-2mm}
\caption{Case study: Hurricane Bonnie. (A) The track detected by {\TE} is colored by wind speed, whereas the track provided by IBTrACS is in black for reference. (B) The track detected by {\tool}, where the default thresholds and human-in-loop thresholds lead to the same result. (C) Selected vector fields with the {\TE} track (1st row) and {\tool} track (2nd row). TC eyes detected from selected time steps are indicated by arrows. These vector fields are located in orange dashed boxes from (A) and (B).  }
    \label{fig:Tails}
    \vspace{-4mm}
\end{figure*}

\begin{figure*}[t]
    \centering
    \includegraphics[width=2.01\columnwidth]{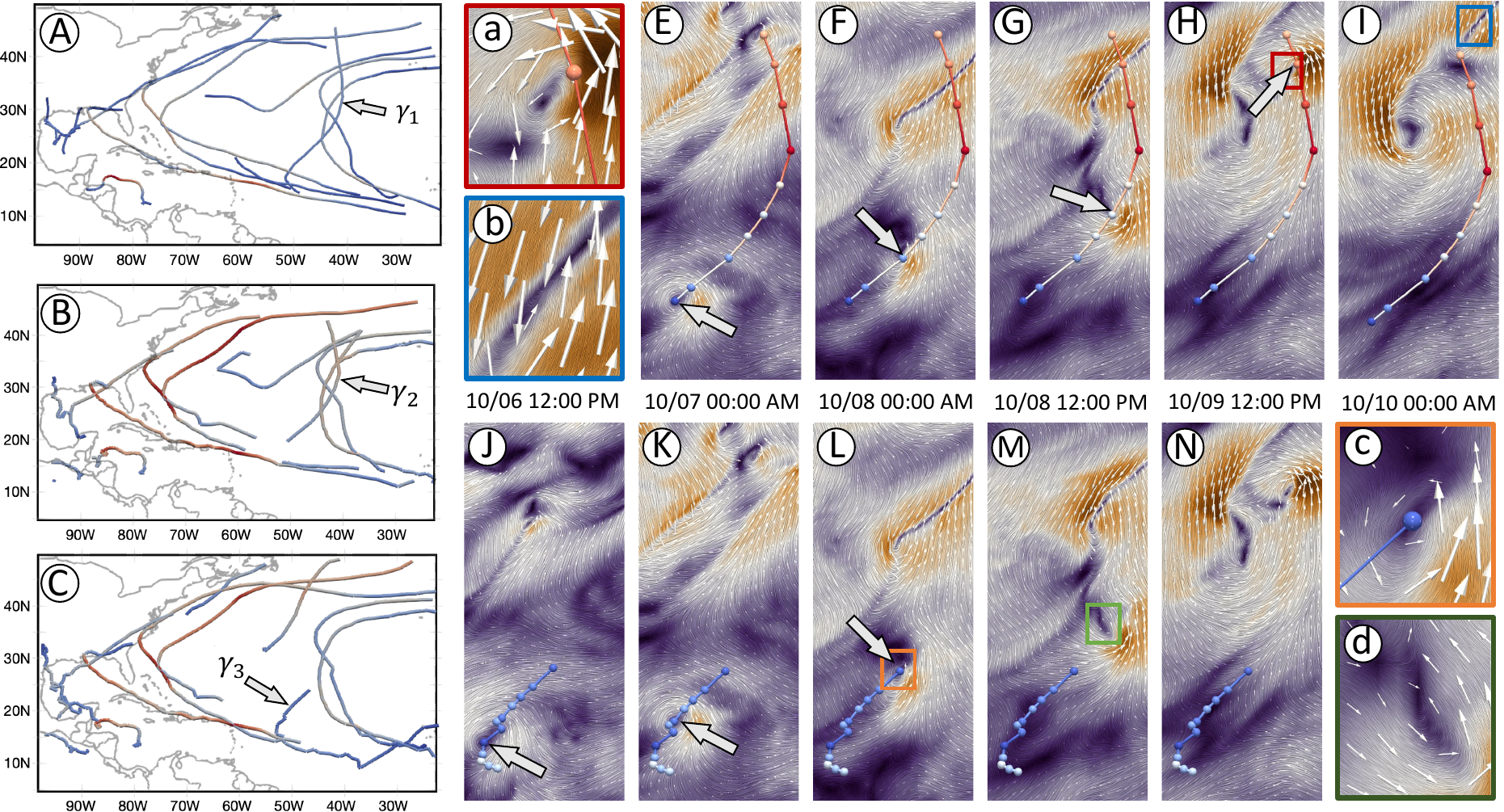}
    \vspace{-2mm}
\caption{Case study: Hurricane Lisa.  1998 annual TCs recorded by IBTrACS (A), detected by {\TE} (B) and {\tool} (C), respectively.
Hurricane Lisa is marked as $\gamma_1$, $\gamma_2$, and $\gamma_3$ in each method.
(E--I) and (J--N) Hurricane Lisa detected by {\TE} and {\tool} with selected vector fields as background. TC eyes detected by each method in corresponding time steps are indicated by arrows. 
(a--d) Zoomed-in views of regions indicated by boxes with the same colors. Tracks in this figure are colored by wind speed.}
    \label{fig:MissedTraj}
    \vspace{-2mm}
\end{figure*}

\subsection{Overview of Results}
We apply {\tool} to the $\EW$ dataset on a cluster with 664 nodes (128GB DDR4 and 36 cores per node). 
We utilize the method of Tricoche~\etal~\cite{TricocheScheuermannHagen2001} to calculate degrees of critical points for the vector field in each time step and parallelize the computation of multilevel robustness with Eden~\cite{SimmermanOsborneHuang2012}, which can schedule and manage a number of tasks on a high-performance computing cluster.

Even though we have obtained the tracking results using {\tool} for all 30 years of data, we demonstrate TC tracking results for 1981, 1990, 2004, and 2009 in~\cref{fig:teaser} to highlight the tracking differences between various datasets and algorithms. 
%The third column shows the results with default thresholds for filters and the fourth column shows the results whose thresholds of filters for each year are tuned by user considering trajectories from IBTrACS as guideline. The TC tracking results using {\TE} are shown on the second column of~\cref{fig:teaser}. 
{\TE} tracks TC eyes as positions with the local minimum SLP, whereas the TC eyes from {\tool} are located where wind speeds are zeros. Therefore, TC tracks from {\TE} and {\tool} do not overlap exactly. 
Both methods consider the area within a two-degree great circle of a detected TC eye to find a local maximum and use it to measure storm intensity~\cite{ZarzyckiUllrich2017,UllrichZarzycki2017}. Tracks in~\cref{fig:teaser,fig:Tails,fig:MissedTraj} are colored by storm intensity~\wrt~this \emph{two-degree maximum wind speed}, referred to as \emph{wind speed} for simplicity. We discuss some preliminary findings based on~\cref{fig:teaser} in the following paragraphs

First, since {\TE} requires both dynamic and thermodynamic variables to meet its specified criteria, it usually cannot detect the beginnings and endings of TC where wind speeds are low. {\tool} can detect such tails as long as the centers of flows can be identified as critical points; see tracks $a$, $b$, and $c$ in~\cref{fig:teaser}. 

Second, {\tool} can detect some discrete storm events that are not captured by \TE, such as tracks $d$ and $e$ in~\cref{fig:teaser}. Track $d$ can be found in IBTrACS, whereas track $e$ cannot. To further investigate these discrepancies, including tracks that cannot be captured by \TE/IBTrACS but are found by {\tool}, and tracks that are in IBTrACS but are undetected by {\tool}, we conducted case studies (reported in \cref{sec:case-study,sec:terminate}). 
%in~\cref{sec:case-study}. Actually, {\tool} may also fail to capture some trajectories that are detected by {\TE} and recorded in IBTrACS. We investigate one missed trajectory with {\tool} in~\cref{sec:terminate}.
 
Third, using fine-tuned filter thresholds for {\tool}, we can get TC tracking results that are more similar to IBTrACS. 
For example, track $f$ is missed by {\tool} with a default threshold for smoothness, since $f$ is severely bent and has a relatively low smoothness value. When, however, we decrease the threshold for $s(\gamma)$ from 0.967 to 0.95, track $f$ is shown in the TC tracking result as indicated by the red arrow in~\cref{fig:teaser}. Similarly, track $e$, which is not recorded in IBTrACS, can be filtered out if we increase the threshold of stability $b(\gamma)$ from 0.029 to 0.03. 
These sensitivities to parameters in the tracking algorithm are, in general, expected \cite{enz2022parallel}. 
This human-in-the-loop option provides users the opportunity of fine-tuning, especially for short-term forecasting or weather-scale studies, as accuracy is more important at this scale. 
At the climate scale and for climate change impacts, users  may use the same thresholds for all TC cases in both historic and future periods, to avoid the impacts of parameter uncertainties. 
 
\subsection{Case Study: Tracking TC during Dissipation}
\label{sec:case-study}

We now investigate a Category 2 hurricane track named Hurricane Bonnie. 
As illustrated in \cref{fig:Tails}, (A) shows the TC tracks from {\TE} (colored by wind speed) and IBTrACS (in black), whereas (B) shows the TC tracks detected by {\tool}.
We observe a much longer tail in (B) with low wind speed compared with (A).
(C) gives the zoomed-in views of tails of TC tracks detected by {\TE} (1st row) and {\tool} (2nd row). Vector fields from different time steps are shown as background. We also highlight the TC eyes detected by each method in corresponding time steps by arrows on (C).
We see TC eyes detected by {\tool} are exactly located where the wind vanishes, whereas the TC eyes from {\TE} are slightly shifted to the outside of the eyes. 
The last two columns of (C) show that {\tool} can track the TC eyes during dissipation because the wind flow is still spinning even if the wind speed is low. 
{\tool} tracks such flow behaviors as sources until the sources disappear.
%\jiali{okay this might be why it's not in real data or TempestExtreem, because in reality, hurricane flow can not be a source, which is a divergence, corresponding good weather in meteorology. if you are comfortable let's remove this part of discussion?} 

%\begin{figure}[h]
%    \centering
%    \includegraphics[width=\columnwidth]{Traj2.pdf}
%    \vspace{-4mm}
%\caption{A trajectory captured by {\tool}, but cannot be founded in \TE/IBTrACS.} \lin{May delete if there is no good story.} \jiali{Agree}
%\jiali{maybe can just mention in conclusion and discussion that {\tool} maybe able to detect new TCs that are missed by real observations somehow.. otherwise may delete the figure as you already have quite a lot of figures.}}}
%    \label{fig:AddTraj}
%    \vspace{-2mm}
%\end{figure}

%\subsection{Case Study: Terminating Tracking When a TC Loses Its Distinct Eye}
\subsection{Case Study: Terminating Tracking} 
%  When a TC Loses Its Distinct Eye
\label{sec:terminate} 

We next investigate a Category 1 hurricane named Hurricane Lisa, which is captured by {\TE} but is partially missed by {\tool}.
\cref{fig:MissedTraj} (A--C) show TCs of 1998, which are recorded in IBTrACS and detected by {\TE} and {\tool} with the $\EEight$ dataset, respectively. 
We focus on the analysis of Hurricane Lisa, marked as $\gamma_1$, $\gamma_2$, and $\gamma_3$ in different methods. 
We note that $\gamma_3$, detected by {\tool}, has a clearly shorter track compared with the other two datasets because the eyes of Hurricane Lisa were detected as centers only up to 10/08/1998 00:00 AM, after which Hurricane Lisa lost its distinct eye and could not be detected by {\tool}; see~\cref{fig:MissedTraj} (M), as well as a zoomed-in view in~\cref{fig:MissedTraj} (d). 

Such a process could be related to TC's extratropical transitions \cite{KnaffLongmoreMolenar2014}, which may occur as Hurricane Lisa moves over cooler water and into areas of stronger wind shear at higher latitudes. During this transition, the cyclone loses its symmetric and distinct eye and, thus, {\tool} terminates its tracking.

In fact, Hurricane Lisa produced distinct eyes again at time step 10/09/1998 at 12:00 PM and 10/10/1998 at 00:00 AM, so {\tool} was able to start the track again. However, because these tracks were short-lived, they were filtered out during the initial feature selection. 
%another eyewall \jiali{you meant eye not eyewall, right? i changed 'eyewall' to 'eye' in above paragraph.} at time step 10/09/1998 12:00 PM, as shown in~\cref{fig:MissedTraj} (H) and (a). But it loses its eyewall \jiali{you meant eye not eyewall, right?} again at 10/10/1998 00:00 AM, as shown in~\cref{fig:MissedTraj} (I) and (b).
%The loss of eyewall \jiali{you meant eye structure not eyewall, right?} makes {\tool} terminate its TC track. (Lin: Can we use it to evaluate the storm simulation model?) \jiali{if Tempest captured that and we don't, it could be the tracking algorithm, if none of us captured it, it could be the data. but i think we've explained it well.}

\subsection{Evaluations: Annual Domain-Averaged Climatology}
\label{sec:AnnualClimatology}
This section and the following present the evaluations of {\tool} using the metrics described in \cref{sec:metrics}. For annual climatology, cumulative statistics are calculated over the entire data period (30 years), which are then normalized to a per-year basis. 
\cref{tab:annual} shows annually averaged statistics for TCs detected by {\tool} and \TE. 

\begin{table}[t]
 \caption{Annually averaged metrics (frequency, TC days,  and latitude of lifetime-maximum intensity) for TCs detected by {\TE} and {\tool} relative to the reference (IBTrACS). }
 \vspace{-2mm}
 \label{tab:annual}
\scriptsize%
  \centering%
  \begin{tabu}{%
  *{4}{c}%
  }
\toprule
&$\overline{count} (\#)$&$\overline{tcd} (days)$&$\overline{lmi}$ ($^{\circ}lat.$)\\
	\midrule   
  IBTrACS &10.74 &79.34&25.65\\
  {\TE} & 8.17&50.01&38.30\\
{\tool} (default) &5.3&52.62&31.55\\
{\tool} (human-in-loop) &7.77&66.43&29.98\\
\bottomrule
  \end{tabu}%
  \vspace{-4mm}
\end{table}
%The top line represents average statistics from IBTrACS for reference.  
%The use of IBTrACS observations as the reference allows us to use these metrics to evaluate TC tracking algorithms. 

According to annual TC count ($\overline{count}$), IBTrACS contains approximately 11 TCs per year within the studied region, whereas {\tool} with default setting produces the least number of TCs. However, when considering annual storm lifetime (\ie, $\overline{tcd}$), {\tool} detects longer storms than {\TE} and is closer to IBTrACS on average. 
This result is consistent with what we observed in our experiments. 
{\tool} can detect TCs not only during their movements but also, at least partially, during their formation and dissipation periods, even if their wind speed is low; see~\cref{fig:Tails} for an example.
Also, since {\TE} requires dynamic and thermodynamic variables to meet specified criteria, TC tracks may break into pieces when some parts of the track do not meet the requirement; see the track indicated by a purple circle from~\cref{fig:teaser} for an example.
Overall, {\tool} produces TC tracks for a longer time than does \TE, whereas {\TE} detects more TC tracks than does {\tool}.

In addition, although {\tool}'s TC tracks have closer hurricane genesis $\overline{lmi}$ \wrt~the observation, both {\tool} and {\TE} show a poleward bias compared with IBTrACS's genesis. %in term of the hurricane genesis $\overline{lmi}$. 
One of the reasons for this bias could be that the reanalysis data including ERA5 are not able to  simulate storm structures near the equator well, according to Knaff~\etal~\cite{KnaffLongmoreMolenar2014}, indicating that the input data to any tracking algorithms is the most important factor when studying TCs.
%As pointed out by Knaff~\etal~\cite{KnaffLongmoreMolenar2014}, the manifestation of reanalysis data (including ERA5) is likely more poorly simulating storm structure closer to the equator. The basis of $\overline{lmi}$ from~\cref{tab:annual} is in agreement with this statement, since TCs detected from the $\EW$ dataset tend to increase in size at higher latitudes. But {\tool} produces TC tracks has closer $\overline{lmi}$ \wrt~the observation compared with \TE. 

%%jiali here
\subsection{Evaluations: Spatial Climatology}
\label{sec:spatialClimatology}
\begin{figure}[t]
    \centering
    \includegraphics[width=\columnwidth]{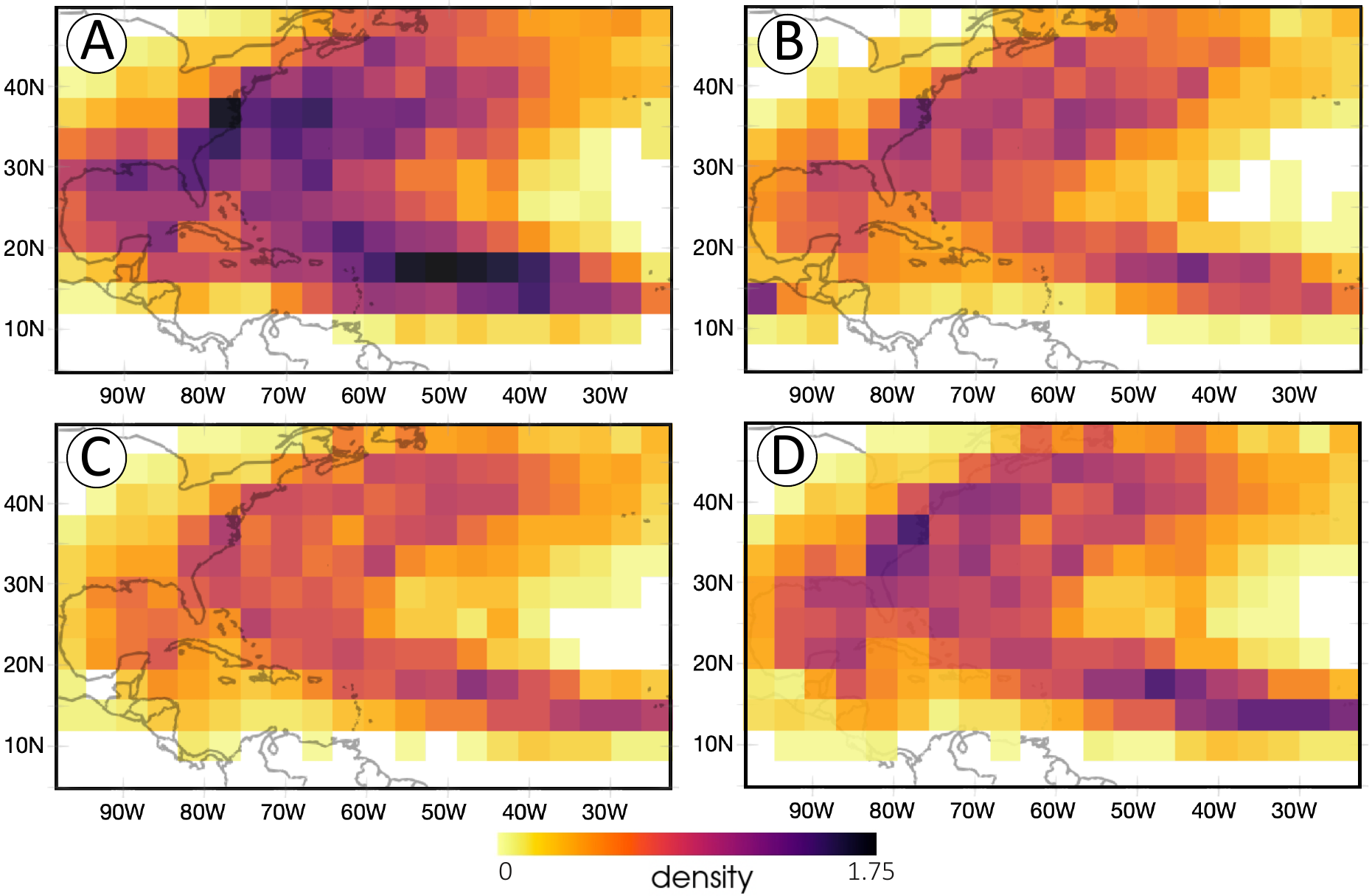}
    \vspace{-6mm}
\caption{Annual track density plots for all TCs detected by IBTrACS (A), {\TE} (B), {\tool} (C, default), and {\tool} (D, human-in-the-loop). Units are 6-hourly TC fixes per $4^\circ \times 4^\circ$ grid box. }
    \label{fig:rxyTrack}
    \vspace{-4mm}
\end{figure}

For spatial pattern climatology, density maps are generated by aggregating occurrences into $4^\circ$ by $4^\circ$ bins.
%, showing how frequent each dataset can produce TCs. 
An example of the spatial track density patterns is shown in~\cref{fig:rxyTrack}. The spatical correlation between IBTrACS and {\tool} (default) is caculated using patterns from~\cref{fig:rxyTrack} (A) and (C). To quantify the similarity between IBTrACS and {\tool} (or TempeExtremes), we calculate Pearson correlation coefficient $r_{xy}$ for total occurrence ($r_{xy,track}$), genesis occurrence ($r_{xy,gen}$), maximum wind speed, and minimum SLP. 
%For instance, to calculate the total occurrence frequency ($r_{xy,track}$) and genesis ($r_{xy,gen}$) frequency correlations, the raw number of hits (per year) are summed in each bin. 
%An example of the spatial track density patterns used to derive the correlations is shown in~\cref{fig:rxyTrack}. 
%After getting the density maps, we can evaluate the spatial patterns between {\tool}/{\TE} and observation using Pearson correlation coefficient described in~\cref{eq:PearsonCor}. 
\cref{tab:spatical} shows the pattern correlation of {\tool}/{\TE} with observations. 
%The first metric is cumulative track density ($r_{xy,track}$). 

Both {\tool} and {\TE} can produce reasonable distributions of storm occurrence when compared with observations ($>0.89$). 
The spatial patterns of genesis ($r_{xy,gen}$) from {\tool} and {\TE} show lower correlations ($<0.64$) with observations. This is not surprising because the initialization and development of a clear eye depend on many other factors such as their 3D evolutions, which are beyond the capability of what our 2D vector fields can represent. In fact, predicting the genesis of TC is one of the scientific challenges in TC research fields, indicating that the currently available data cannot  capture TC genesis and that new frameworks of such calculations are needed \cite{yang2021hurricane}. 
%The storm first appears when the data assimilation system is employed to generate new initial conditions for the forecast, suggesting that its development was not easy to predict from a forecast run initialized 12 hours earlier. Thus data preparation is likely one factor in the inability to extract a clear center when the storm is a tropical depression. Tropical depressions are also not well-organized systems, the development of a clear eye is dependent on their 3D evolution beyond our study of 2D vector fields, which make {\tool} also has low consistency with observation in detecting the genesis of TCs.
%The last two correlations query spatial patterns of different intensity measures, $r_{xy,u10}$ and $r_{xy,slp}$ measure the spatial correlation as measured by maximum u10 and minimum sea level pressure, respectively, over each grid box over the duration of the data period.  
Both {\tool} and {\TE} are able to produce  storm strength similar to that of observations ($\ge 0.89$) in terms of maximum $u_{10}$ and minimum sea level pressure. 

\begin{table}[t]
\caption{Spatial correlation for TCs detected by {\TE} and {\tool} with the reference (IBTrACS). 
%The TC activity is binned into $4^{\circ}\times 4^{\circ}$ boxes prior to correlation calculation. Spatial climatology includes spatial pattern of cumulative track density ($r_{xy,track}$), genesis ($r_{xy,gen}$), maximum $u10$ ($r_{xy,u10}$), and minimum SLP ($r_{xy,spl}$). 
}
\vspace{-2mm}
 \label{tab:spatical}
\scriptsize%
  \centering%
  \begin{tabu}{%
  *{5}{c}%
  }
\toprule
$4^{\circ}\times 4^{\circ}$ & $r_{xy,track}$ & $r_{xy,gen}$ & $r_{xy,u10}$ & $r_{xy,slp}$\\ [0.5ex] 
\midrule   
 IBTrACS & 1.00 & 1.00 & 1.00 & 1.00  \\ 
 {\TE} & 0.892 & 0.638 & 0.920 &0.889 \\
{\tool} (default) & 0.897 & 0.569 & 0.925 & 0.901 \\
{\tool} (human-in-loop) & 0.914 & 0.634 & 0.933 & 0.920 \\
\bottomrule
  \end{tabu}%
  \vspace{-4mm}
\end{table}

\section{Conclusion and Discussion}
\label{sec:conclusion}

We introduce a physics-informed TC tracking framework, {\tool}, that utilizes tools from vector field topology. Based only on a 2D wind vector field, {\tool} is able to produce results comparable to (and sometimes even better than) those obtained with a widely used TC tracking algorithm---{\TE}---while requiring far less input data.
Although {\tool} does not consider the air temperature field (e.g., the warm cores) of TCs, the symmetric eye structures of TCs allow {\tool} to detect and track them. \myedit{Furthermore, our framework may be used in uncertainty visualization to understand  uncertainty due to different model physics or setup, or feature comparison of geoscientific data with space and time dimensions.}

\myedit{{\tool} has a number of limitations. First, our robustness-based framework is very useful for hurricane tracking, since hurricanes have symmetric structures and hurricane tracks are one of the most important factors in assessing their risks. However, once the symmetric structures (i.e., the eyes) are weakened or disappear with the cyclones moving to higher latitudes, {\tool} does not consider them as TCs anymore. This is because the cyclone at higher latitudes may get their energy from one or more front systems dividing warm air from the south and cold air front the north, see~\cref{fig:MissedTraj}. Such frontal systems lead to asymmetric structures, which are not detected by {\tool}. 
In general, our technique may not be suitable for asymmetric feature tracking such as extratropical cyclones.  
Second, to obtain optimal tracking results, we may need to fine-tune the parameters for each single event. 
%This is not always required if we study the climate change impacts on TC features by tracking a current against a future time period, where same parameters would ensure a comparison between the two time periods. 
Third, the current framework only considers the near surface (10 meters) winds. However, higher altitude winds (dozens to hundreds of meters) are also big concerns when it comes to real-world applications such as wind energy. This is left for future work.}

\myedit{From an application perspective, to the best of our knowledge, it is an open challenge to incorporate 3D data in the study of TCs, based on domain scientist feedback. 
Because adding more variables may reduce the overall efficiencies of TROPHY yet may not guarantee a better performance.
From an algorithmic perspective, it may be possible to extend {\tool} to utilize 3D data for TC tracking. First, we may use horizontal layers of a 3D vector field along the vertical direction. Second, we may utilize a third variable called the vertical motion (i.e., upward and downward), in addition to zonal wind (U) and meridional (V) wind. Expanding {\tool} to either of these directions could be useful in better detecting, tracking and understanding hurricanes such as their genesis, intensification, and landfall (important factors to be considered for risk assessment). The robustness framework has been extended previously to study 2D symmetric tensor fields~\cite{WangHotz2017,JankowaiWangHotz2019} and critical points in 3D vector fields~\cite{SkrabaRosenWang2016}. It may be feasible to integrate the robustness framework for 3D critical points in {\tool}. However, there are more complex features that need to be considered in 3D, such as vortex regions or vortex core lines. 
We would need to develop theoretical foundations to quantify their robustness first before establishing their physical interpretability in studying TCs.}
%i moved this to the first paragraph when talking about strength. \myedit{Furthermore, our framework may be used in uncertainty visualization to understand  uncertainty due to different model physics or setup, or feature comparison of geoscientific data with space and time dimensions.}

%Moreover, although this study tracks TC eyes with symmetric features, the framework can take into account more dynamic and thermodynamic variables to enable more functionalities in studying asymmetric features, such as extratropical cyclones, which can bring storm surges, high winds, and heavy precipitation.
%\myedit{For example, \tool can detect extratropical cyclones when their winds do some rotations, \eg, during the stage of ''occluded cyclone''. Adding the gradient of temperature can differentiate such detected extratropical cyclones with TCs during TCs analysis and visualization. For the stages that cannot be detected by \tool,  \eg, during the stage of ''open wave'', identifying stationary fronts created by cold air and warm air may help to study asymmetric features.}

\acknowledgments{
This study is supported by the Wind Energy Technologies Office of the DOE. Argonne National Laboratory is a US Department of Energy laboratory managed by UChicago Argonne, LLC, under contract DE-AC02-06CH11357. This research was also supported by DOE DE-SC0022753, DOE DE-SC0023193, DOE DE-SC0021015, NSF IIS 2145499, NSF IIS 1910733, and DOE DE-AC02-06CH11357.
}

\clearpage
\appendix
\section{\myedit{An Example of Merge Tree}}
\label{sec:exampleMT}

We now give an example of an augmented merge tree that is constructed from the $\EW$ dataset.
First, we define a scalar field $f_0:\Xspace \to \Rspace$ by assigning the vector magnitude to each point $x \in \Xspace$, that is, $f_0(x) = ||f(x)||_2$; see~\cref{fig:CalRob}(A) and (B), which visualize the vector field $f$ and scalar field $f_0$.
Second, we track the merging behavior of components that contain critical points of $f$. For example, the component containing $x_1$ and $x_2$ merges with the component containing $x_3$ at $r=0.84$ and forms $C_3$ in $\Xspace_{0.84}$, which is represented by the purple and green region in~\cref{fig:CalRob} (D).
Third, we augment the merge tree with the degrees of critical points (on leaves) and the degrees of components (on internal nodes). For example, component $C_1$ in~\cref{fig:CalRob} (D) contains critical points $x_1$ and $x_2$, whose degrees are $+1$ and $-1$, respectively. The degree of $C_1$ is $\mydeg(x_1)+ \mydeg(x_2)=0$, i.e., $\mydeg(C_1)=0$. The augmented merge tree of~\cref{fig:CalRob} (A) is shown in~\cref{fig:CalRob} (E).

\section{\myedit{An Example of Robustness Calculation}}
\label{sec:exampleRob}

\begin{figure}[h]
    \centering
    \includegraphics[width=\columnwidth]{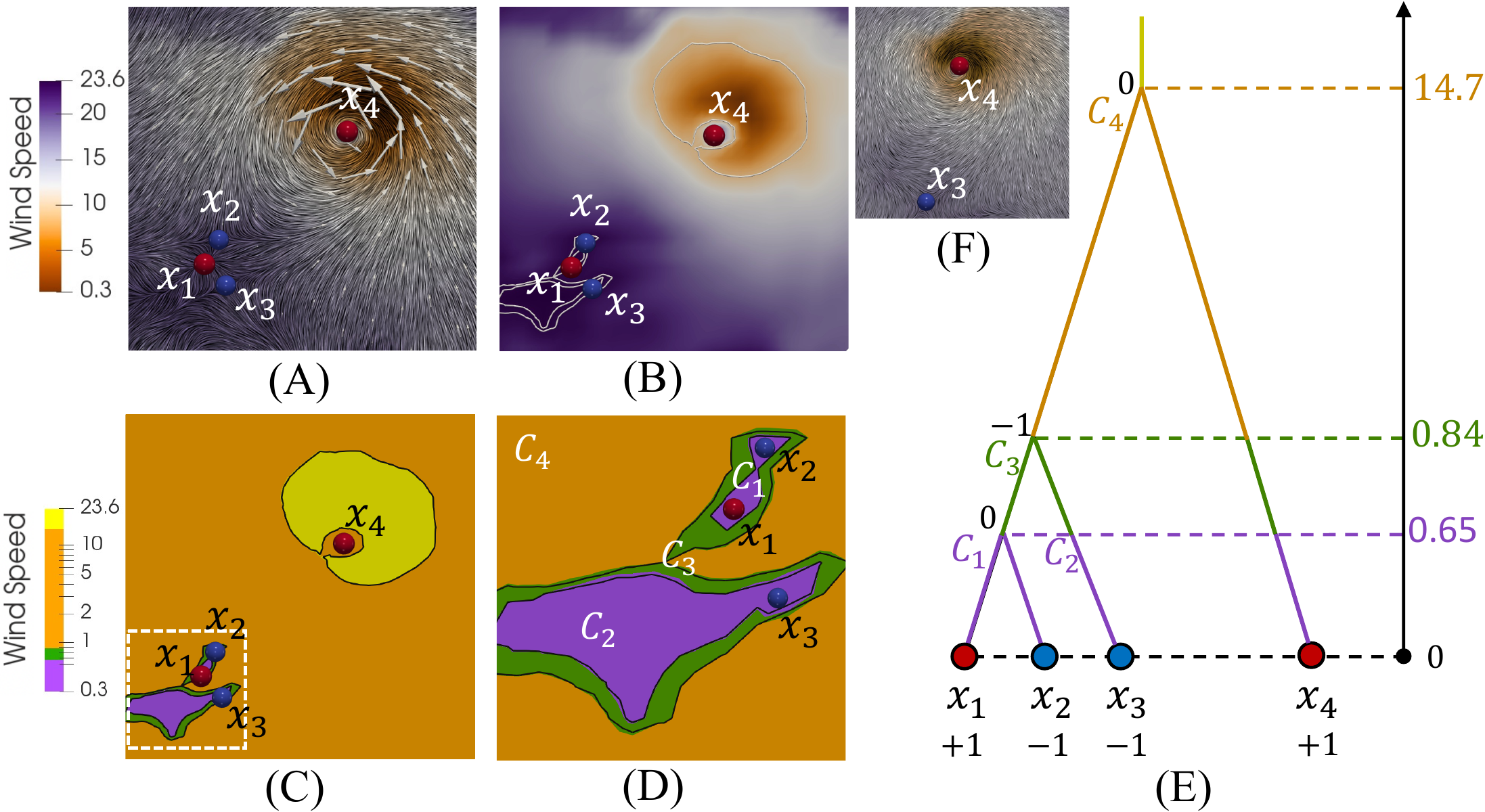}
    \vspace{-6mm}
\caption{Computing robustness with an augmented merge tree. (A) A 2D vector field $f$ and (B) its corresponding scalar field $f_0$. (C) Multiple sublevel sets of $f_0$. (D) A zoomed-in view of the white box in (C). (E) The augmented merge tree. (F) The vector field one time step behind (A). Sources/sinks/centers are in red, and saddles are in blue.}
    \label{fig:CalRob}
    \vspace{-4mm}
\end{figure}

Using the example in~\cref{fig:CalRob} (A)-(E), we now show how to calculate robustness with an augmented merge tree.
As pointed out in~\cref{sec:classicRobustness}, the robustness of a critical point can be calculated as the function value of its lowest zero-degree ancestor in the augmented merge tree.
The robustness of $x_1$ and $x_2$ is 0.65, whereas the robustness of $x_3$ and $x_4$ is 14.7.
Intuitively, for the example in~\cref{fig:CalRob} (A)-(E), it is easier for $x_1$ and $x_2$ to be canceled with each other than $x_3$ and $x_4$, since they have much lower robustness values. In~\cref{fig:CalRob} (F), we give the vector field from the same dataset but one time step (6 hours) behind the vector field of ~\cref{fig:CalRob} (A). We see $x_1$ and $x_2$ disappear, whereas $x_3$ and $x_4$ remain.

\section{Details on Dataset and Methods}
\label{sec:data}

We demonstrate the performance of {\tool} using 30-year (1981--2010) near-surface wind vector field from the ECMWF Reanalysis v5 (ERA5). It is produced by the Copernicus Climate Change Service (C3S)~\cite{C3S}. ERA5 provides hourly estimates of the global climate information with a spatial grid resolution of 30 km. 
%In this paper, we focuses on the analysis of tropical cyclones/storms with the time period 1981-2010. 
Since tropical cyclones/storms usually occur during June and October, we limit our dataset with a time window from June $1$ to October $31$ every year at standard synoptic reporting times (0000, 0600, 1200, and 1800 UTC).
A rectangle region on the Atlantic Ocean ($5$ N$^\circ$ to $49.5$ N$^\circ$ and $98$ W$^\circ$ to $18$ W$^\circ$) is selected.
We utilize 10-meter zonal and meridional wind speed as the 2D vector field, since in the near-surface the hurricane core represents a region of strong convergence and associated vertical motion.
We annotate this 30-year dataset as the $\EW$ dataset. We also mark the one-year subset data from the $\EW$ dataset as $\EY$; for example, \cref{fig:filterBeforeMR} uses the $\EFour$ dataset. 
%\begin{figure}[h]
%    \centering
%    \includegraphics[width=\columnwidth]{freqCyclones.png}
%    \vspace{-4mm}
%\caption{The average occurrences of tropical cyclones/storms per month between 1981-2010 from IBTrACS.}
%    \label{fig:TCOccurence}
%    \vspace{-2mm}
%\end{figure}

%\para{Method}.

We use the International Best Track Archive for Climate Stewardship (IBTrACS~\cite{KnappKrukLevinson2010} version 4) observations as the reference.
IBTrACS is compiled from quality-controlled records from various forecasting centers. 
In this paper, we select TCs reported by World Meteorological Organization (WMO) official forecast centers. Again, only tropical cyclones/storms within the region of the $\EW$ dataset are visualized.
For comparison purposes, we include the TC tracking results of the \TE~software package~\cite{UllrichZarzycki2017} applied to the $\EW$ dataset with the parameter setting following~\cite{ZarzyckiUllrich2017}. 
%We use these tracking results to make a comparison between {\tool} and the state-of-the-art method using the same dataset.

\end{document}